# Training Set Augmentation and Biology-Aware Harmonization Improve Radiomic Models for Lung Cancer Prediction in Indeterminate Nodules

Claire Huchthausen,[1,2] Menglin Shi,[1,3] Gabriel L.A. de Sousa,[1] James Larner,[4] Einsley Janowski,[4] Jonathan Colen,[5] and Krishni Wijesooriya[1,4]*

*Corresponding author: Krishni Wijesooriya, Department of Radiation Oncology, University of Virginia School of Medicine, 200 Jeanette Lancaster Way, Charlottesville, VA 22903. E-mail: kw5wx@virginia.edu

[1] Department of Physics, University of Virginia, Charlottesville VA 22904 USA

[2] Present affiliation: Department of Physics, Massachusetts Institute of Technology, Cambridge MA 02139 USA

[3] Present affiliation: Department of Biomedical Engineering, Northwestern University, Evanston IL 60208 USA

[4] Department of Radiation Oncology, University of Virginia, Charlottesville VA 22904 USA

[5]Old Dominion University, Norfolk VA 23529 USA

**Abstract**

CT radiomics-based machine learning has potential to predict lung cancer in pulmonary nodules (PNs) earlier than standard-of-care methods. Low malignancy rates in early-development PNs and variable image acquisition hinder development of radiomic models for diagnosing these PNs. To address these challenges, we augmented training using later-development PNs and harmonized for acquisition effects. We examine early-development benign and malignant PNs (n=106) below the sensitivity of standard-of-care diagnosis. Classifiers predicting malignancy performed near chance when trained on ComBat-harmonized radiomic features from only early-development PNs. We then augmented training with later-development benign and malignant PNs (n=225). We evaluated whether harmonization must incorporate biology that impacts acquisition effects in added training data. To correct variability from four acquisition protocols, we compared: 1) biology-unaware harmonization, 2) harmonizing with a covariate distinguishing early-development, later-development benign, later-development malignant datasets, 3) harmonizing each dataset separately. Models trained using augmentation, but biology-unaware harmonization, failed to improve consistently. Augmented training data harmonized with a covariate (ROC-AUC 0.74 [0.69-0.79]) or separately (ROC-AUC 0.71 [0.66-0.77]) yielded higher test ROC-AUC (Delong, $p \leq 0.05$) and PR-AUC (Wilcoxon, $p \leq 0.05$). In a proof-of-principle methodological study, we demonstrate with a small single-center dataset that combining radiomic features from later-development benign and malignant PNs requires biology-aware harmonization.

**Keywords:** lung cancer, diagnosis, radiomics, harmonization, data augmentation, machine learning.

# Introduction

Lung cancer causes one in five cancer deaths worldwide [1]. Patients whose lung cancer is detected at an early stage when the cancer is localized exhibit 61% five-year relative survival, compared to 7% in those whose cancer is undetected until distant metastasis [2]. Even before metastasis, larger tumors lead to higher treatment-related toxicity [3]. Adults at risk for lung cancer due to smoking history or age range undergo annual screening with computed tomography (CT) scans, where physicians identify abnormal growths, pulmonary nodules (PNs), and mark these as suspicious for cancer if they meet certain criteria [4], primarily size and growth rate. Small and stable PNs early in their development are typically classified as benign or indeterminate. Some may represent slow-growing cancer to which standard-of-care methods are not sensitive. Conversely, growth can be caused by benign inflammatory processes [5]. Other visual cues like smoothness, irregularity, and spiculation are subject to physicians' interpretation [6]. A definitive diagnosis requires follow-up scans over several months, plus Positron Emission Tomography (PET) scans or biopsy.

Medical images contain information indiscernible to the naked eye [6],[7],[8] extractable using radiomic features, curated statistical and spatial interrelationships between selected voxels in an image [7]. Machine learning (ML) models can harness radiomic features to predict clinical outcomes [6],[8]. Radiomics-based ML could potentially distinguish benign from malignant PNs before the standard-of-care is sensitive to their differences. However, there are two major barriers to creating radiomics-based models to classify indeterminate PNs. First, radiomic features are sensitive to image acquisition protocols [9] which vary across and within clinics. This non-biological variability makes it difficult to meaningfully combine data or reproduce results [10]. Second, there is a low cancer occurrence rate in PNs that are very small or early in their development—the PNs that can be below the sensitivity of standard methods. For example, when Lung-RADS criteria were retrospectively applied to scans from the National Lung Screening Trial, only 0.39% of patients with PNs classified as benign or probably benign at baseline would be diagnosed with cancer after follow-up [11].

To address this class imbalance to improve performance, ML training data may be augmented with radiomic features from confirmed benign and malignant PNs later in their development. Later-development PNs have more pronounced malignant or benign signals and may train ML to better recognize subtle manifestations of the same signals in early-development PNs. However, including these PNs in training may exacerbate the challenge of radiomic features' dependence on acquisition protocols, especially when biological differences in later-development benign and malignant PNs cause their scans to be affected differently by acquisition protocols. For example, contrast enhancement (CE), administered during scans for cases with clinical concern [12], generally accumulates more in tumors than healthy tissue [13]. Both CE's frequency (administered more/less often) and effect (accumulates more/less) create a quasi-systematic difference between later-development malignant and benign PNs that their radiomic features might inherit when calculated from routine imaging.

To mitigate effects of variable image acquisition protocols and improve model generalizability, ComBat ("Combine Batches") harmonization can learn corrective transformations to remove acquisition-related bias [14],[15] from radiomic feature distributions [16],[17],[18]. In CT-based radiomics for lung cancer, ComBat has been applied mostly in pipelines predicting outcomes like recurrence [19], development-free survival [20], and overall survival [21]. These predictive problems use data from only malignant PNs. Lung cancer diagnosis requires data from both benign and malignant PNs. Some studies implemented ComBat when combining radiomic data from benign and malignant tissues, but most corrected for multi-center overall effects rather than variation in specific acquisition protocols [22], [23], [24]. The challenge of harmonizing radiomic features from benign and malignant tissues resembles the challenge of harmonizing features from tissues of different origin. Previous studies consider methods of applying ComBat to harmonize radiomic feature data from tissues of different origin (breast tumor and healthy liver tissue) when acquisition protocols vary, including biology-aware methods of harmonizing with tissue type as a covariate [25] and harmonizing different tissue types separately [25],[26]. This study will investigate whether radiomic data from benign and malignant lung tissue similarly require biology-aware harmonization when accounting for variable acquisition protocols.

To develop radiomics-based models to classify indeterminate PNs from routine imaging, we augmented training with later-development benign and malignant PNs and accounted for non-standardized acquisition using ComBat harmonization. To remove radiomic features' dependence on four acquisition protocols, we compared three methods of harmonizing augmented training sets: 1) harmonizing all training data collectively in the standard manner unaware of biology, 2) harmonizing with a covariate to distinguish early-development data, later-development benign augmentation data, and later-development malignant augmentation data, and 3) harmonizing these three datasets separately. We compared models trained on augmented data harmonized using these methods by their performance classifying PNs below the sensitivity of standard-of-care lung cancer detection methods.

# Results

**Baseline radiomic model for lung cancer prediction**

We collected a dataset containing clinical chest CT scans of PNs early in their development, where their small size and subtle features place them below the cancer detection threshold of standard-of-care methods. For the purposes of this study, all PNs that were interpreted at time of scan as benign, probably benign, or indeterminate are considered below the cancer detection threshold of standard-of-care diagnosis methods. We term these early-development PNs. Each PN was scanned multiple times separated by interims of months, and each scan had a corresponding time-of-scan interpretation.

PNs imaged in these early-development scans received follow-up. At the end of follow-up, some PNs were diagnosed as benign and some PNs were diagnosed as malignant (Methods). We refer to this as a PN's endpoint diagnosis, which may differ from its time-of-scan diagnoses (for examples see Supplementary Fig. S1). The early-development scans of PNs with an endpoint malignant diagnosis are the PNs for which sensitivity could be improved. Overall, the early-development dataset had only 29.2% incidence of cancer (Table 1). Throughout this study, we evaluate model performance using held-out test data drawn only from this early-development dataset.

We developed a predictive pipeline using this data, shown schematically in Fig. 1. PNs were segmented in the images using semi-automated contouring software. Scans were preprocessed (Methods) and 107 standard radiomic features were extracted for each scan of a PN using the Python library PyRadiomics (Fig. 1a) [8]. This early-development dataset was randomly split into five folds, while ensuring no patients overlap between training and test sets, repeated 10 times (50 trials; Fig. 1b). This controls for longitudinal progression of a PN, not examined here. Each scan is classified without referencing other scans of that PN, mirroring how lung cancer screening begins at an arbitrary point relative to PN development.

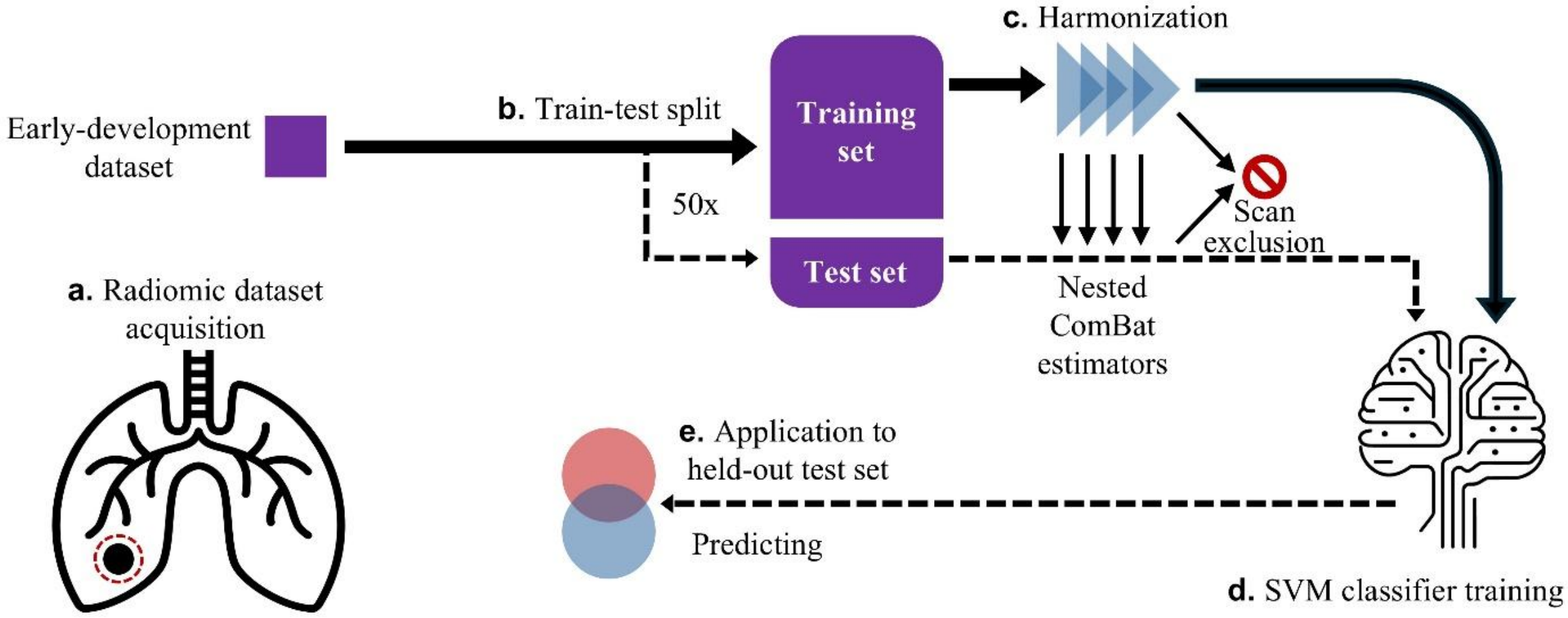

**Figure 1. The pipeline for our baseline lung cancer prediction model.** The solid line represents the training set and the dashed line represents the test set for a given trial. An arrow pointing from the training set to the test set indicates that a corresponding result from training set processing is used to process the test set.

The dataset included scans acquired using non-standardized protocols, including kilovoltage peak (KVP), focal spots, scanner manufacturer, and CE. For each training set, we used Optimized Permutation Nested ComBat (OPNCB), which applies ComBat to remove biases from multiple protocols sequentially [27], to correct for variability in these four protocols (Fig. 1c).

To predict scan endpoint diagnosis from harmonized radiomic features, we trained an L1-regularized linear support vector machine (SVM) classifier. From all radiomic features in the training set, the regularization selected a mean=10.3 [9.4-11.2] and median=10.0 features per trial (n=50; brackets indicate the 95% confidence interval here and henceforth). The SVM was trained to classify scans by endpoint diagnosis, benign or malignant (Fig. 1d).

The trained ComBat estimators and trained SVM were applied to the held-out test set for each trial (Fig. 1e). ComBat was applied to the test set by matching a given test sample's acquisition protocols to the estimators already trained for those protocols; no other information from the test set was used. This model trained on only early-development data achieved an average test ROC-AUC=0.58 [0.52-0.64] (n=50) and PR-AUC=0.44 [0.37-0.51]. This performance is near chance, motivating our hypothesis that augmented training data may improve classification of early-development PNs.

### Training set augmentation with radiomic features from later-development PNs

To improve performance by teaching models strong malignant/benign phenotypes, we added radiomic feature data from benign and malignant PNs later in their development (n=225 scans, Table 1) to the early-development training data. For the purposes of this study, PNs interpreted at time of scan as suspicious or malignant are later-development; these are above the cancer detection threshold of standard-of-care methods. Later-development PNs have 550% larger volumes than early-development scans on average and many of these scans were the last taken before PET or biopsy. We refer to later-development benign and malignant scans as augmentation datasets because they augment early-development training data but are not used to test performance. We refer to training sets containing data from both early- and later-development datasets as augmented training sets. Figure 2 visualizes our dataset definitions and Supplementary Figure S1 follows three representative PNs to further clarify classification. The malignant dataset holds later-development scans of PNs that were later confirmed as malignant using PET or biopsy.

The benign dataset holds later-development scans of PNs that were later confirmed as benign using wedge resection or biopsy. Later-development datasets include follow-up scans of PNs that have previous scans in the early-development dataset. To ensure independence in each trial, if a patient was assigned to the test set, no scans from that patient, including all later-development follow-up scans, were included in the training set. With augmentation datasets, cancer occurrence increased from 29.2% to 66.5% in our total data pool (Table 1), resolving the challenge of low malignancy rates. We added these augmentation datasets to the original early-development training sets in all trials and trained classifiers following the pipeline in Fig. 3.

Later-development datasets were also highly variable in our acquisition protocols of interest (Supplementary Table S1). Note that, prompted by their size, later-development PNs are 172% more frequently scanned using CE. To evaluate radiomic features' acquisition dependence in each augmented training set, we developed a procedure using the Kruskal-Wallis test [17] (Methods). A significant result (adjusted $p \leq 0.05$) indicated a feature was acquisition-dependent. Table 2 summarizes feature dependence on acquisition protocols and Fig. 4 shows $p$-values for acquisition dependence of each feature and acquisition protocol from a sample trial. Without harmonization, 15.0% [13.9-16.0%] of radiomic features were acquisition-independent (n=50 trials, Fig. 4 top row). To mitigate effects of acquisition variation within each trial, we applied OPNCB to learn corrective transformations from the entire augmented training set (hereafter: collective harmonization, Fig. 3 bottom left). Features remaining acquisition dependent were excluded from further analysis in each trial (Fig. 3c*). However, collective harmonization lowered the proportion of acquisition-independent features to 13.2% [12.3-14.2%] (n=50 trials; Wilcoxon signed-rank test, $p$=0.008). Collective harmonization increased CE dependence in later-development benign data, and total acquisition dependence, compared to unharmonized data (Table 2).

**Table 1. Contents of the datasets.** Patient and PN counts do not add to the column totals because augmentation datasets include follow-up scans of PNs in the early-development dataset.

| | Patients | | Pulmonary nodules | | Scans | |
|---|---|---|---|---|---|---|
| | **Count** | **% Cancer** | **Count** | **% Cancer** | **Count** | **% Cancer** |
| Early-development | 39 | 38.5 | 54 | 29.6 | 106 | 29.2 |
| Augmentation benign | 21 | 0.0 | 28 | 0.0 | 36 | 0.0 |
| Augmentation malignant | 58 | 100.0 | 59 | 100.0 | 189 | 100.0 |
| Total | 100 | 60.0 | 120 | 50.8 | 331 | 66.5 |

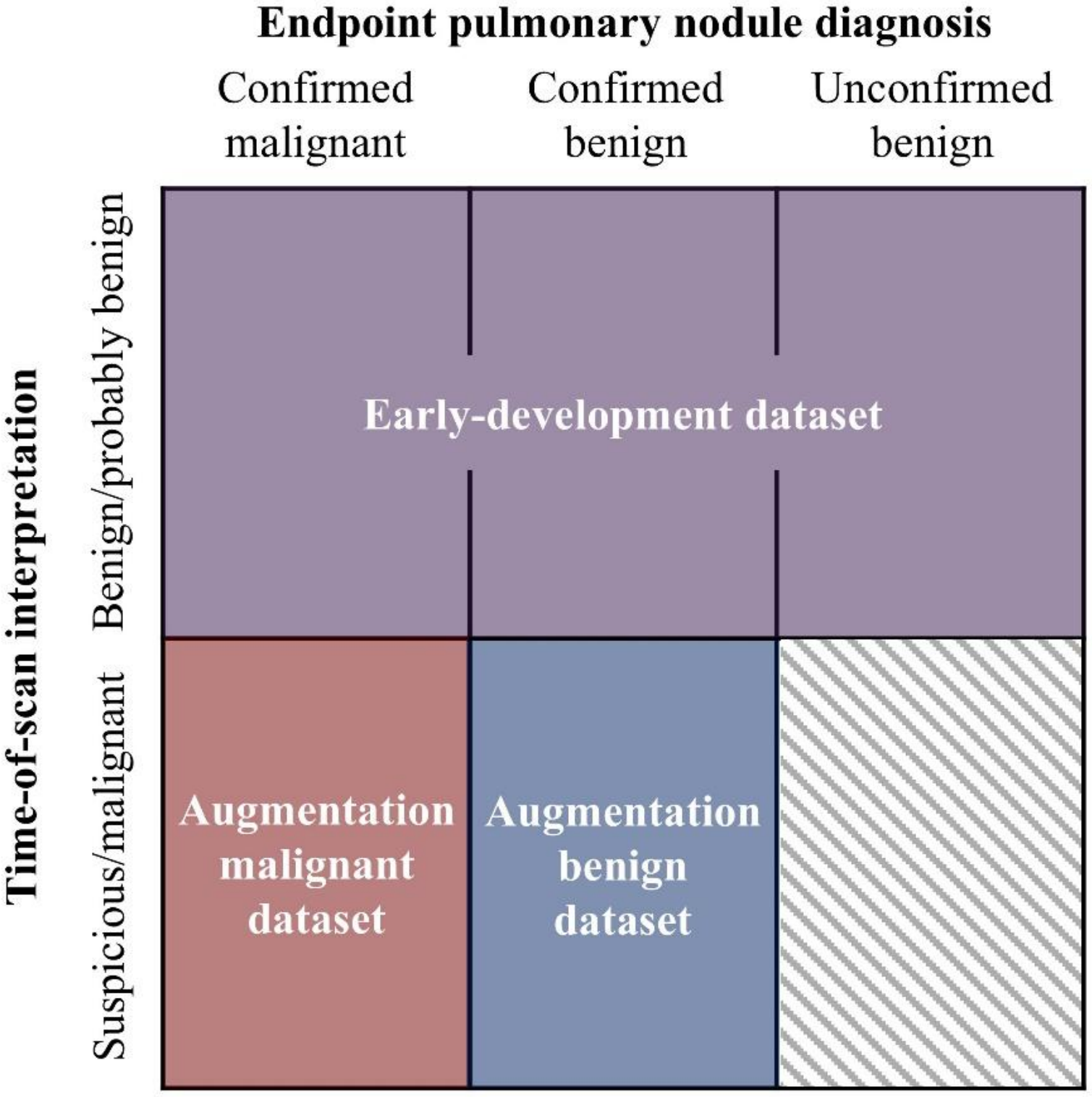

**Figure 2. Definitions of how scans were classified into datasets.** Each cell represents a category of scans defined by the intersection of the axes labels, with colors indicating the dataset to which that category was assigned. The greyed-out region indicates a category which does not exist because all PNs interpreted as suspicious/malignant in our datasets received follow-up to confirm diagnosis.

**Comparing harmonization methods to remove acquisition dependence in augmented training sets**

Collective harmonization assumed each acquisition protocol biased a radiomic feature in the same way across all three datasets, yet this assumption may be invalid due to biological and acquisitional differences between early-development, later-development benign, and later-development malignant datasets. Radiomic features in later-development benign and malignant datasets may have different distributions due to their biological differences independent of acquisition. Specific protocols are used to acquire scans in different datasets at different frequencies (Supplementary Table S1) due to clinical selection or chance, further influencing the feature distributions. Using ComBat with a covariate—an added corrective protocol learned for a specific dataset—is designed to account for acquisition-based differences in distributions, while still assuming an acquisition protocol has the same effect on scans in biologically distinct datasets [25]. A transformation applied to scans in different datasets has the same slope but a net shift in intercept [25].

When a protocol has different effects on scans in biologically distinct datasets, corresponding corrective transformations may have to differ between datasets in both intercept and slope. Because ComBat estimates acquisition effects from all input data, datasets with different acquisition effects (additive or multiplicative) should be harmonized separately [25]. For example, different tissue types (breast tumor and healthy liver tissue) require distinct ComBat transformations to correct for acquisition variability [25] due to their biological differences.

CE exemplifies a difference in both the means and acquisition effects between our later-development benign and malignant augmentation datasets. Later-development scans administered with CE due to clinical concern are more likely to be malignant (Supplementary Table S1), creating a difference in mean for some radiomic features, and contrast medium is expected to accumulate more in tumors than healthy tissues [25], creating a difference in acquisition effect for some radiomic features. Collectively harmonized features exhibited frequent acquisition dependence due to CE, indicating collective harmonization does not handle these differences well.

Thus, to account for baseline and acquisition differences between biologically distinct PNs, we compared other methods of applying the OPNCB algorithm: 1) harmonizing with a covariate to distinguish early-development, late-development benign, and late-development malignant datasets (hereafter: covariate harmonization, Fig. 3 bottom center), or 2) harmonizing early-development, late-development benign, and late-development malignant datasets separately (hereafter: separate harmonization, Fig. 3 bottom right).

When all augmentation data was added to the training set (Fig. 3c*), the proportion of acquisition-independent features defined by Kruskal-Wallis averaged 40.5% [38.8-42.2%] (n=50) with covariate harmonization, and 93.5% [90.9-96.1%] (n=50) with separate harmonization. The frequency of each feature's net acquisition dependence after each method—how often that feature's distribution was dependent on at least one acquisition protocol—is in Supplementary Table S2. When comparing metrics for the methods, adjusted *p*-values were calculated using Holm-Bonferroni. Covariate harmonization increased the proportion of acquisition-independent features compared to collective harmonization (Wilcoxon, adjusted $p$=2.2e-09, $n$=50). Separate harmonization increased the proportion of acquisition-independent features compared to collective harmonization (Wilcoxon, adjusted $p$=2.2e-09, $n$=50) and covariate harmonization (Wilcoxon, adjusted $p$=2.2e-09, $n$=50).

**Table 2. Frequency of feature dependence on acquisition protocols.** Frequency percentage measures the rate of acquisition dependence, defined as a significant Kruskal-Wallis test result, per acquisition protocol across all features and trials (n = 50 × 107 = 5350). Percentages do not sum to the given totals due to redundant failure.

| | Unharmonized (%) | | Collective harmonization (%) | | Covariate harmonization (%) | | Separate harmonization (%) | |
|---|---|---|---|---|---|---|---|---|
| | Benign | Malignant | Benign | Malignant | Benign | Malignant | Benign | Malignant |
| CE | 0.0 | 78.5 | 45.2 | 78.8 | 24.9 | 27.6 | 0.0 | 3.7 |
| Focal spot size | 43.0 | 41.1 | 7.9 | 1.0 | 23.1 | 19.5 | 0.0 | 6.5 |
| KVP | 0.0 | 0.0 | 0.0 | 0.4 | 0.3 | 2.7 | 0.0 | 0.0 |
| Manufacturer | 0.0 | 40.2 | 0.0 | 0.9 | 1.3 | 9.6 | 0.0 | 0.0 |
| Dataset total | 43.0 | 82.2 | 52.4 | 78.8 | 42.2 | 38.9 | 0.0 | 6.5 |
| Method total | 85.0 | | 86.8 | | 59.5 | | 6.5 | |

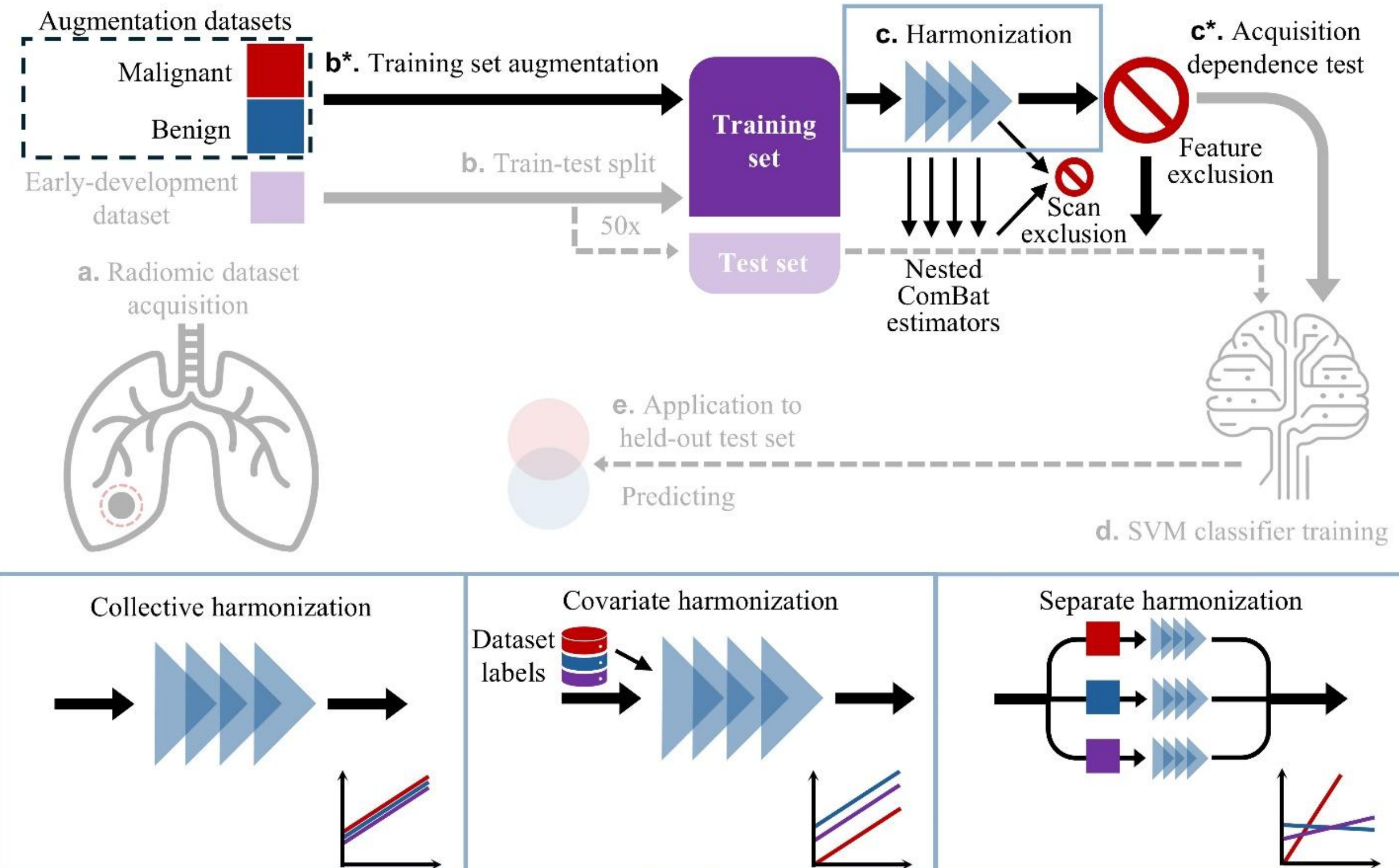

**Figure 3. The pipeline for lung cancer prediction models with augmented training sets.** Pipeline elements with reduced opacity are procedures that remain unchanged from Fig. 1. Asterisks (*) indicate new steps in the pipeline as compared to Fig. 1. The solid line represents the training set and the dashed line represents the test set for a given trial. An arrow pointing from the training set to the test set indicates that a corresponding result from training set processing is used to process the test set. Figure 3c represents one of the three harmonization methods expanded schematically at the bottom of the figure. Within the expansions (bottom), line graphs illustrate the type of corrective transformations in feature space that each method calculates.

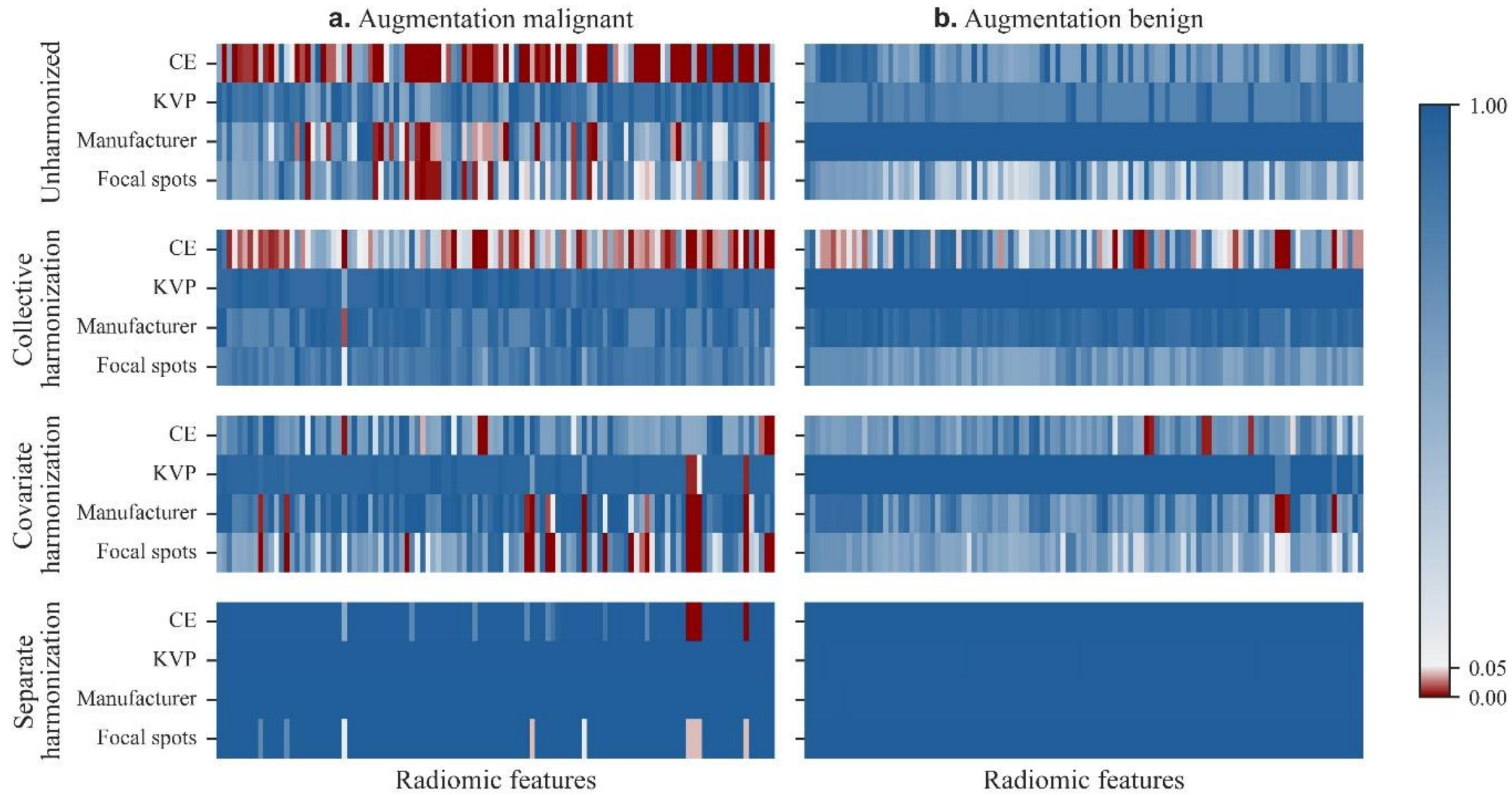

**Figure 4. Acquisition dependence of radiomic features defined by the Kruskal-Wallis test in an example training set.** *P*-values ≤0.05, indicating dependence of a feature (107 total, x-axes) on an acquisition protocol (y-axes), are in red. Values for each feature were adjusted across protocols using Benjamini-Hochberg to control the false discovery rate. More red cells in a row show a greater frequency of dependence on the corresponding acquisition protocol. a. Acquisition dependence within scans from the augmentation malignant dataset. b. Acquisition dependence within scans from the augmentation benign dataset.

### Impact of training set augmentation and harmonization method on classification

To observe how performance scales with additional training data, we constructed a learning curve by varying the fraction of the augmentation datasets added to each training set. For each augmentation level of each training set, SVMs were trained on features corrected using collective, covariate, or separate harmonization, creating three versions of a model for each augmented training set. We refer to these models as collective, covariate, and separate harmonization models depending on the harmonization method applied prior to training.

While biology-aware covariate and separate harmonization distinguish benign and malignant PNs in the augmentation datasets, they make no distinction when harmonizing the early-development dataset. Applying trained ComBat estimators to the test set, drawn only from the early-development dataset, thus uses no foreknowledge of whether each PN is benign or malignant.

With biology-unaware collective harmonization, performance improved at small augmentation levels but declined at augmentation levels greater than 25% (Fig. 5a,c left). Both biology-aware harmonization methods enabled consistent gains as more training data was added. ROC-AUC and PR-AUC scaled logarithmically for covariate (Fig. 5a,c center) and separate (Fig. 5a,c right) harmonization models.

When the training set included 100% of augmentation data, biology-aware separate and covariate harmonization methods yielded significantly better performance than biology-unaware collective harmonization, at the scan level (Table 3a) and at the PN level (Table 3b). We include sensitivity at 30% specificity and specificity at 90% sensitivity when evaluating model performance because these thresholds approximate the sensitivity and specificity of Lung-RADS on intermediate-size PNs at baseline, as found by a retrospective study of the National Lung Screening Trial [28]. Our early-development dataset includes both intermediate-sized PNs and much smaller PNs; using these thresholds compares standard-of-care performance on an easier task to radiomic models' performance on our harder task. By our study design, standard-of-care methods had a sensitivity of 0 on all test samples when seen in the clinic.

**Table 3. Performance metrics for the three harmonization models.** 100% augmentation. Weighted accuracy uses balanced weighting by class (endpoint benign or malignant diagnosis). *Significant (Wilcoxon, adjusted $p<0.05$) improvement over collective harmonization. **Significant (Wilcoxon, adjusted $p<0.05$) improvement over all other harmonization methods.

| | **Collective Harmonization** | | **Covariate Harmonization** | | **Separate Harmonization** | |
|---|---|---|---|---|---|---|
| | **Mean, %** | **95% CI, %** | **Mean, %** | **Mean, %** | **95% CI, %** | **Mean, %** |
| ***3a. Performance per scan*** | | | | | | |
| ROC AUC | 57.9 | [51.8, 64.0] | 74.0* | [69.0, 78.9] | 71.4* | [66.0, 76.7] |
| PR AUC | 43.9 | [36.7, 51.0] | 60.7** | [53.5, 67.9] | 55.5* | [47.9, 63.2] |
| Weighted accuracy | 58.7 | [54.1, 63.2] | 66.2* | [62.2, 70.1] | 63.3 | [59.6, 67.1] |
| Specificity at 90% sensitivity | 25.3 | [17.6, 33.0] | 42.2* | [34.1, 50.3] | 41.1* | [32.0, 50.1] |
| Sensitivity at 30% specificity | 73.8 | [66.0, 81.6] | 89.5* | [83.2, 95.9] | 88.9* | [83.6, 94.3] |
| Sensitivity at 90% specificity | 22 | [14.0, 30.0] | 43.5* | [35.5, 51.5] | 34.9* | [26.2, 43.6] |
| ***3b. Performance per PN, averaged over scans*** | | | | | | |
| ROC AUC | 61.1 | [54.6, 67.7] | 73.6* | [67.4, 79.8] | 74.0* | [68.7, 79.4] |
| PR AUC | 47.5 | [39.0, 55.9] | 63.6* | [55.4, 71.8] | 60.6* | [52.5, 68.7] |
| Weighted accuracy | 62.7 | [58.1, 67.4] | 67.9* | [63.4, 72.4] | 61.8 | [58.0, 65.6] |
| Specificity at 90% sensitivity | 33.8 | [24.9, 42.7] | 51.6* | [42.3, 61.0] | 51.9* | [42.3, 61.5] |
| Sensitivity at 30% specificity | 78.2 | [70.4, 86.1] | 88.7* | [82.6, 94.9] | 87.8* | [81.2, 94.4] |
| Sensitivity at 90% specificity | 25.5 | [16.4, 34.6] | 42.8* | [33.0, 52.7] | 38.5* | [29.2, 47.8] |

Because lung cancer screening begins at an arbitrary point in PN development, a valuable radiomic model should increase sensitivity at any isolated point in care. Thus we evaluate model performance on individual test scans in our analysis henceforth. By the Delong test, covariate harmonization models outperformed collective harmonization models (adjusted $p \leq 0.05$ in 16.7% trials, n=48) and separate harmonization models outperformed collective harmonization models (adjusted $p \leq 0.05$ in 14.3% trials, n=49). Separate and covariate harmonization did not differ in ROC-AUC, but covariate harmonization achieved higher PR-AUC (Wilcoxon, p=0.019, n=50). ROC-AUC and PR-AUC did not differ significantly between scanner manufacturers except PR-AUC for covariate harmonization, suggesting separate harmonization may be more robust to scanner imbalance.

Calibration curves and decision curve analysis are in Supplementary Fig. S2. Biology-aware harmonization models were better calibrated than collective harmonization models (Wilcoxon, p = 3.8e-04, 2.9e-13). Separate harmonization models have higher net benefit than both "treat all" and "treat none" strategies at decision thresholds within [0.33, 0.52]; a "treat none" strategy was utilized for all test samples when encountered in the clinic.

**Predictive features in augmented training sets**

We again analyze trials where 100% of available augmentation data were added to the training set. From acquisition-independent harmonized features, a mean=7.7 [6.9-8.5] and median=7.0 features were selected per trial for collective harmonization, mean=17.06 [15.9-18.2] and median=17.0 for covariate harmonization, mean=17.4 [16.3-18.6] and median=17.0 for separate harmonization (n=50). The most frequently selected features varied by harmonization method (complete frequencies in Supplementary Table S3): flatness for collective harmonization; GLRLM long run high gray level emphasis, sphericity, GLCM cluster shade, flatness for covariate harmonization; and sphericity, GLCM cluster shade, GLCM informational measure of correlation 2, GLDM large dependence high gray level emphasis, GLSZM small area low gray level emphasis, NGTDM strength, flatness, and kurtosis for separate harmonization. A previous study used flatness to classify benign and malignant PNs [29]. Radiomic texture features--GLCM, GLDM, GLSZM, NGTDM—are assumed to reflect biological characteristics like tumor heterogeneity, demonstrated indirectly [30]. Pearson correlation [31] assesses feature selection stability over cross-validation in Supplementary Table S4.

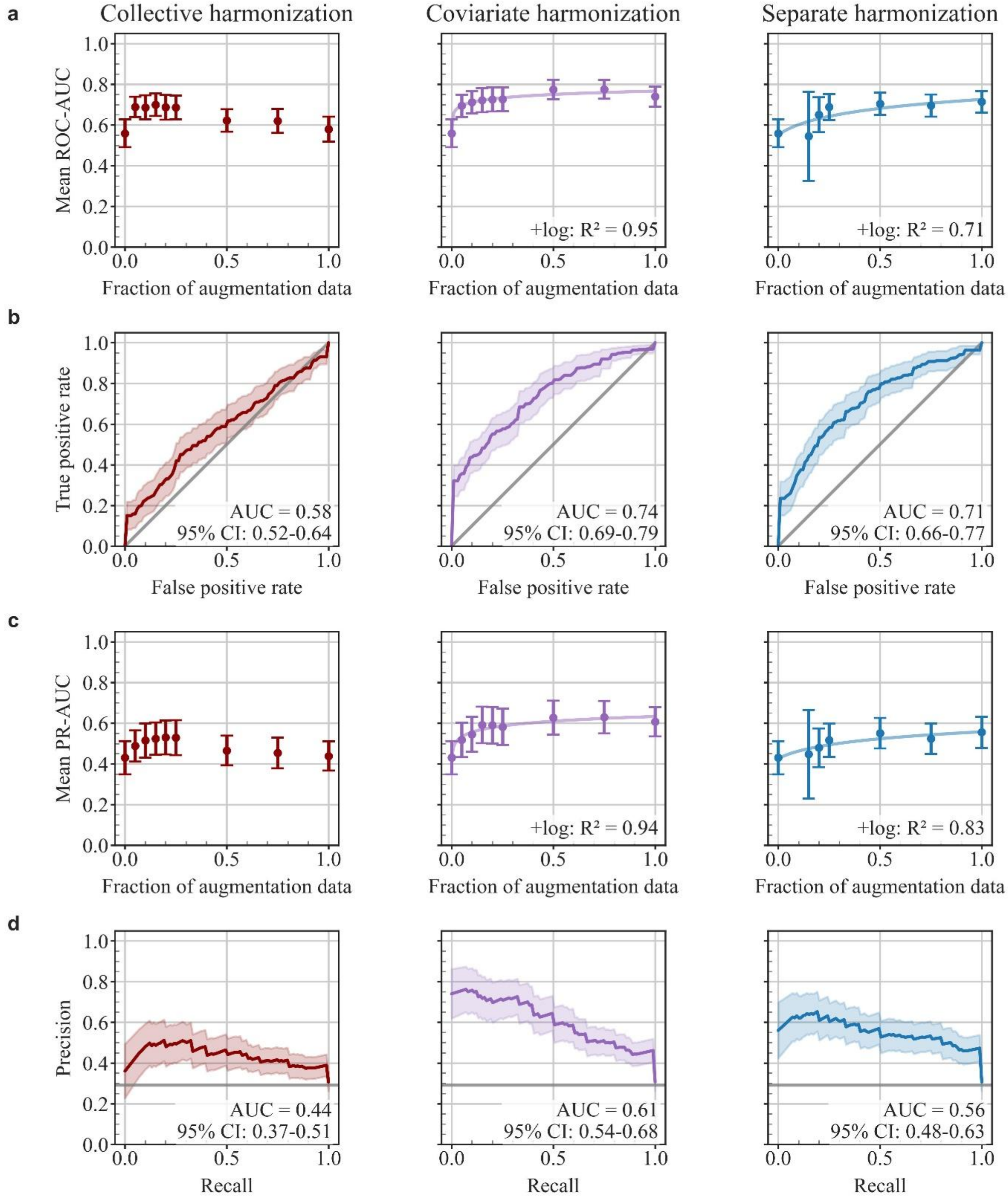

**Figure 5. Average performance of the three harmonization models.** Results are reported at the scan level. a. (c.) The ROC-AUC (PR-AUC) descriptive learning curves for models with augmented training sets obtained by varying the fraction of augmentation data included in the training set. The leftmost point in each plot represents the baseline un-augmented model. Error bars show the 95% confidence interval over trials where a model could be trained. Trial numbers *n* for each point, from left to right—collective harmonization: *n*=50 for all, covariate harmonization: *n*=50 for all, separate harmonization: *n*=[50, 10, 30, 45, 50, 50, 50]. Separate harmonization failed for 5% and 10% augmentation due to small sample size, so these points are omitted. b. (d.) Receiver operating (precision-recall) curves

for models trained using 100% of available augmentation data, averaged over all trials, n=50 for each. Shaded regions show the 95% confidence interval. Gray lines indicate chance-level performance (ROC-AUC: 0.5, PR-AUC: 0.29).

# Discussion

The potential of radiomics to improve sensitivity in lung cancer prediction is determined by how well a radiomics-based model can handle real-world data challenges. Such a model must overcome an inherent sample class imbalance and non-standardized image acquisition. In this proof-of-principle study, we addressed class imbalance by augmenting training data with later-development benign and malignant PNs and addressed acquisition variability using ComBat harmonization. Correcting acquisition effects is essential because models trained on acquisition-dependent radiomic features do not generalize well to new sites [32],[25] or rare acquisition protocols. Models trained without harmonization, using acquisition-dependent data, may not appear to perform badly. But patterns learned by the model may be an artifact of acquisition—an acquisition protocol's effect or frequency of use—that differ distinctively between benign and malignant PNs. These effects prevent a model from generalizing and are precisely what harmonization aims to remove.

Harmonization becomes complicated when augmenting training data with later-development PNs because radiomic features from biologically distinct PNs—early-development, later-development benign, later-development malignant—may need distinct corrections to compensate for acquisition variability. E.g., CE may have a large effect on feature distributions from later-development malignant PNs, being used more frequently and having enhanced permeability and retention in these PNs, but a smaller effect on features from later-development benign PNs (Fig. 4 top row). Consistent with this, later-development malignant PNs required significantly larger feature-space displacements during harmonization than early-development or later-development benign PNs to become acquisition-independent (Supplementary Fig. S3). While we focus on differences across later-development datasets, early-development PNs may show feature distributions distinct from later-development PNs, one example being PN volume.

Incorrectly harmonized radiomic features may mislead a model rather than improve performance. Indeed, when harmonization was biology-unaware, we found performance decreased with greater augmentation. It is possible that low augmentation boosted performance similarly to noise injection [33]. Thus, we evaluated biology-aware harmonization methods. While collective harmonization applied identical corrections to all datasets, separate and covariate harmonization learned corrections specific to each biologically distinct dataset (Fig. 3). Covariate harmonization accounted for dataset-specific feature distributions (from differences in biology or differences in acquisition protocol frequencies due to clinical selection or chance) and separate harmonization further accounted for dataset-specific acquisition effects [25]. Separate and covariate harmonization recovered acquisition-independent distributions in benign and malignant later-development training data for significantly more features than collective harmonization, suggesting radiomic features of later-development benign and malignant PNs need distinct corrections. Separate harmonization removed acquisition dependence from significantly more features than covariate harmonization, suggesting acquisition protocols have different effects on benign and malignant PNs for many radiomic features (Supplementary Table S2). But separate harmonization's aggressive corrections risk removing biologically meaningful variation, and its corrections are limited by smaller sample size [25]. These may account for separate harmonization's slightly lower performance than covariate harmonization models.

Conversely, does distinct treatment in harmonization introduce information that could artificially improve performance? Orlhac et al. addressed this question when combining data from different tumor stages, demonstrating that even covariate harmonization does not artificially add information that could influence a classifier [25]. Consistent with this conclusion, feature displacement in covariate-harmonized test samples did not differ from other methods (Wilcoxon, p=0.83; Supplementary Fig. S3) and UMAPs (e.g. Supplementary Fig. S4) qualitatively showed no new clustering in covariate-harmonized data.

Impacts of the acquisition protocols we harmonized have been studied [9], [34], [35], [36]. Our datasets are heterogeneous in several acquisition protocols (Table S1) for which we could not harmonize due to small sample sizes for protocols. Inter-rater variability in segmentations is another limitation of our datasets, mitigated using semi-automatic contouring [37],[38] and intensity thresholding [39]. Kruskal-Wallis analyses show moderate rater effect size in collective and covariate harmonization data, but minimal effect size in separate harmonization data, and very few harmonized features with significant rater-dependence per trial. This requires attention in later work, but exploratory analyses indicate inter-rater variability does not systematically influence model performance in this study. Because unseen batch effects—acquisition protocols and rater variability—are present in trials of each harmonization method, they do not invalidate our comparison of harmonization methods.

This study preserved real-world challenges in lung cancer screening. Our datasets were gathered from a clinical setting and had high acquisition variability, providing a realistic testbed for harmonization methods. Endpoint diagnosis is unknown when harmonizing and classifying early-development scans, as in true diagnosis. Early-development scans were the same seen in true lung cancer screening. As such, some uncertainty in these scans' endpoint diagnosis is inherited from the clinical process. The contents of the early-development dataset are detailed in Supplementary Table S5. 44% of early-development benign PNs (Supplementary Table S5) never required biopsy or PET/CT. For such presumed-benign PNs, we used strict inclusion criteria (Methods) to ensure stability over time. The concern remains that some are nonetheless malignant and skew model predictions. However, Kruskal-Wallis analyses showed that feature distributions from presumed-benign scans are not more malignant-like. Further, biology-aware harmonization produced higher performance both when classifying only confirmed-label test scans (Supplementary Table S6) and when including presumed-benign scans (Table 3a), so we consider potential label noise to have minimal impact on our conclusions regarding these methods. PNs of this type are important to include because they figure in most lung cancer screening scans. Early-development scans with malignant endpoint diagnoses (Supplementary Table S5) also have uncertain ML labels— they may not yet exhibit malignant characteristics—but these scans allow us to evaluate whether radiomic models can improve the standard-of-care sensitivity to lung cancer.

In this methodological study, we found that radiomic features from later-development PNs, when harmonized in a biology-aware manner, can augment the training of ML models and improve their ability to predict cancer in early-development PNs in a small, single-center dataset. Because this is a single-center study, findings are not yet generalizable; external multicenter validation of the value of these methods on a large dataset of early- and later-development PNs is vital. Supplementary Fig. S5 provides preliminary pseudo-external validation by holding out scans by an acquisition protocol, but this leads to sample sizes that are too small to draw reliable conclusions. Harmonization is central to the presented methodology and multicenter data is its gold-standard test. Multicenter data used for validation should have significant overlap in acquisition protocols to allow for harmonization. Inter-rater robustness of segmentations should be quantified, and fully automated segmentation should be considered to reduce segmentation labor. Harmonizing for more acquisition protocols should be explored, and clear conclusions are needed regarding when covariate versus separate harmonization should be used for biologically distinct images.

To meaningfully combine radiomic features from early-development and later-development PNs and harness this information for a tool that could someday improve on standard-of-care diagnosis, we must first establish methods to remove acquisition dependence from these different data types. Until evaluation on a larger dataset, we recommend that radiomics researchers harmonize radiomic features from early-development, later-development benign, and later-development malignant PNs either separately or with a covariate. Studying harmonization methods is an important first step towards radiomics as a real-world quantitative tool for improving the standard-of-care sensitivity to lung cancer, which may support future efforts to identify malignant tumors earlier and reduce unnecessary imaging and invasive procedures.

# Methods

**Radiomic dataset acquisition**

This retrospective study was approved by the UVA Institutional Review Board (IRB) (approval number: 301447), and due to its retrospective nature, informed consent was waived by the UVA IRB. All methods were performed in accordance with the relevant guidelines and regulations. We acquired chest CT scans of PNs from the University of Virginia hospital. Physician annotations and comments on scans were available for all PNs. If resection, biopsy, or PET/CT were performed, these results were available. Some patients had multiple PNs and some PNs had multiple follow-up scans separated by interims of months (Table 1).

All PNs in the early-development dataset were interpreted as benign, probably benign, or indeterminate at time of scan. Keyword indicators of this physician interpretation were, in order of the priority with which they were used to classify scans: benign (Lung-RADS category 2), probably benign (Lung-RADS category 3), inflammatory/infectious, indeterminate, 12-month follow-up (without verbal indication of suspicion), 6-month follow-up (without verbal indication of suspicion).

After follow-up, PNs from scans in the early-development dataset either received a malignant diagnosis, confirmed by biopsy or PET/CT, a benign diagnosis, confirmed by biopsy or wedge resection, or showed no indicators of becoming malignant. These last PNs are presumed benign, having never required biopsy or PET/CT. Because they were not confirmed by these diagnostic methods, presumed benign PNs were only included when at least three scans were performed for that patient following the incidence of the PN, i.e. at least four scans total, to demonstrate stability over time. We additionally require that no subsequent screening scan up to November 2025 could have an unresolved suspicious indication.

This study's evaluations of model performance use individual scans from the early-development dataset as the statistical unit, excepting Table 3b. If a PN has multiple scans, longitudinal information per PN and true scan order is disregarded because the model is not trained on any scans of the same PN.

All scans in our augmentation datasets were interpreted as suspicious or malignant at time of scan. Keyword indicators of this physician interpretation were, in order of priority: malignant, known cancer, very suspicious (Lung-RADS category 4B), suspicious (Lung-RADS category 4A), possibility for neoplasm, recommendation for PET/CT or biopsy (without verbal indication of indeterminate or alternative diagnosis).

The malignant augmentation dataset holds scans of PNs that were later confirmed as malignant by biopsy or PET/CT. The benign augmentation dataset holds scans of PNs that were later confirmed as benign by biopsy or wedge resection.

CT images for all datasets were acquired in DICOM format. CE information was noted from scan protocol comments and all other acquisition protocol information was extracted from the DICOM header files. To acquire segmentations, images were displayed with the lung viewing setting. Following physician annotations, each region of interest corresponding to a PN in the CT images was segmented using semi-automated contouring tools in the software Velocity AI v. 4.1.4 (Varian, A Siemens Healthineers Company, Palo Alto, CA, USA) by one of three raters. Although inter-rater segmentation differences affect radiomic features, intensity thresholding (discussed below) and semi-automated contouring [37],[38] mitigates this effect.

Pixel data and segmentations were extracted from the DICOM files. Because radiomic texture features require voxels to be rotationally invariant [7] the PyRadiomics feature extractor [8] was used to resample images and masks to 1.0 $mm^3$ voxels. Lanczos interpolation was used for the images for greatest reproducibility [40] and nearest-neighbor interpolation was used for segmentation masks. To exclude air (-1000 HU) and bone (cortical: >1000 HU, trabecular: 300-800 HU) voxels, while retaining contrast-enhanced tissue (up to 500 HU), and to improve reproducibility [39], voxels with intensities outside -700 to 500 Hounsfield Units were excluded from calculations of non-shape features. Other radiomic studies of PNs have used significantly narrower thresholds to analyze only malignant tissue [39], but we chose a less conservative threshold because we also analyzed benign tissue and very small PNs. Our threshold is also inclusive of calcifications—these are often identified in practice as >200 HU—which information could aid in classifying PNs as malignant or benign [41]. All other PyRadiomics feature extractor settings were left at default values. Bin-width was fixed at 25 gray values. All PyRadiomics operations were performed with v 3.1.0. A total of 107 standard features from the PyRadiomics library were extracted from all regions of interest.

**Formation of training and test sets for cross-validation of lung cancer prediction models**

In our baseline pipeline, the early-development dataset was split into five folds using a randomized Stratified Group K-fold (scikit-learn), ensuring no patients overlap between training and test sets while preserving the percentage of samples in each class as much as possible per fold, repeated 10 times with a trial-dependent random seed (see code in GitHub), for a total of 50 unique training and test sets.

In our pipeline with training set augmentation, we began with the same splits of the early-development dataset as used to validate the baseline model (50 unique training and test sets), and augmentation data were added to these training sets. For a given trial, scans were not used to augment the training set if the same patient appeared in that trial's test set; such augmentation scans were unavailable to that trial. For each early-development train-test split, we varied the fraction of available augmentation data included in the corresponding training set: 5%, 10%, 15%, 20%, 25%, 50%, 75%, and 100% (8 unique fractions). Fractions were stratified by class and drawn from the available augmentation data using a trial-dependent random seed (see code in GitHub). Training set augmentation was cumulative. This gave a total of 50×8=400 unique augmented training sets corresponding to the 50 test sets.

Repeated runs of Stratified Group K-fold cross-validation controls for patient- and PN-level clustering, ensuring the model generalizes without relying on this clustering for test set predictions. Stratified cross-validation was chosen over bootstrapping due to its lower variance and to guarantee all early-development scans are represented in the test sets [42].

**Harmonization of training and test sets**

This study used OPNCB procedure developed by Horng et al. to correct for acquisition protocols. In OPNCB, harmonization is performed for each acquisition protocol iteratively in an order which removes acquisition dependency from the most features according to the Anderson-Darling test, as compared to other permutations [27]. Because ComBat is a data-driven method, no phantom imaging is required for this standardization [17].

Correction parameters were calculated for a given fold by applying OPNCB to a training set. No information from the test set influenced the calculated corrections. We studied three methods of applying OPNCB to a training set. In collective harmonization, OPNCB was applied to the complete training set, where extracted radiomic features and corresponding scan acquisition information were the inputs. In separate harmonization, the training set was separated by dataset before applying OPNCB to these subsets individually. In covariate harmonization, OPNCB was applied to the complete training set, and a dataset label for each scan was supplied as an additional input using the categorical_cols argument.

OPNCB was performed on training sets with return_estimates=True, and the trained ComBat estimators were applied to the corresponding test sets using the function neuroComBatFromTraining (release 0.2.12) being developed by Fortin et al. [14],[15]. The estimators appropriate to the acquisition protocols of a given test sample were applied in the order determined for the training set by the OPNCB algorithm. For separate harmonization, estimators applied to the test set were those learned for the early-development training scans alone. All other default arguments were retained.

ComBat fails if fewer than three training samples were acquired with the same protocol, e.g., if only two scans were acquired using a certain focal spot size. In these rare cases, these scans were excluded from the trial. We developed an automated process to perform scan exclusion based on these criteria. Training sample exclusion counts are included in Supplementary Table S7 and more detailed per-trial information for training sample exclusion is in our GitHub repository (commit tag def1b96). No early-development training scans were excluded from a trial and no test scans were excluded from a trial because there was sufficient overlap in acquisition protocols.

In our baseline pipeline, OPNCB was applied using collective harmonization. In our pipeline with training set augmentation, for each of our augmented training sets, we applied OPNCB using our three examined methods in parallel, leading to 3 versions of a given augmented training set and corresponding test set for each fold.

**Evaluation of acquisition dependence of radiomic features**

In our pipeline with training set augmentation, to evaluate whether harmonization removed acquisition dependence from augmented training sets, we used the Kruskal-Wallis test on training data to compare feature distributions grouped by instances of acquisition protocols [17]. For example, feature distributions from CE scans were compared to those from non-CE scans. The test was run separately on the benign and malignant augmentation training data, because correct feature distributions should be consistent across acquisition within benign PNs or within malignant PNs, though benign and malignant distributions may differ. The test was not performed on the early-development training data because of this additional source of variability. For a given radiomic feature, if the test was significant (adjusted $p \leq 0.05$) for either benign or malignant samples, that feature was deemed acquisition dependent. No information from the test set influenced the determination of acquisition dependence and subsequent feature exclusion. This procedure was applied to all 107 features.

To validate this procedure, we applied it to randomized splits of unharmonized scans that were acquired with the same protocols of interest—i.e. “acquisition-independent.” Our datasets contained 11 such uniform groups, ranging in size from 7 to 79 scans. These uniform groups contained both benign and malignant samples. The procedure defined an average 98.4% [98.1-98.7%] (n=55) of features to be acquisition-independent as expected. This supports Kruskal-Wallis as a reasonable test to define acquisition dependence. Variability in unaccounted-for acquisition protocols most likely caused nonzero deviation from normal distributions. Some additional acquisition protocols for which this study did not harmonize are included in Supplementary Table S1.

**Classifier training**

A linear SVM classifier whose target was PN malignancy (class 0: endpoint benign diagnosis, class 1: endpoint malignant diagnosis) was fitted to acquisition-independent features in the training set using LinearSVC from sklearn with random_state=42. We chose a linear SVM for this study because it has few hyperparameters and its behavior is well understood, where a more complex classifier could obscure our analysis of harmonization methodology. Radiomic features were standardized to have zero mean and unit variance using the StandardScaler function from scikit-learn before the classifier was applied. The scaler was fitted using training data only. A linear kernel was chosen due to our small test set and to avoid overfitting. We enforced sparsity of the model using L1 regularization to further improve generalization and better model real-world applicability. Features were considered to be selected based on coefficients calculated by the SVM during the training by using the SelectFromModel function from sklearn with a threshold of 1e-5. No information from the test set influenced feature selection. If harmonization failed an SVM was not trained for that trial (i.e., for separate harmonization at low augmentation fractions).

As a control for studying the impact of harmonization methods and for reproducibility, we present our primary results from an SVM with fixed regularization protocol C=0.1. When C was instead tuned for each trial via nested cross-validation, performing a grid search over C=1e-03 to C=100, PR-AUC did not differ significantly from fixed-C models (Wilcoxon, adjusted $p \leq 0.05$) and ROC-AUC did not differ significantly from fixed-C models (Delong test, adjusted $p \leq 0.05$) except for one covariate harmonization trial. We present ROC-AUC and PR-AUC results from tuned models in Supplementary Fig. S6.

All classifier results are reported with a scan as the statistical unit, to compare with standard-of-care methods at any single point of care. AUC values were calculated using the decision scores from LinearSVC. Prediction probabilities for use in calibration curves and decision curve analysis were calculated using isotonic regression, implemented with CalibratedClassifierCV from sklearn and five folds of cross-validation.

The trained SVM classifiers were applied to the held-out harmonized test set corresponding to each training set.

**Statistical analyses**

The Kruskal-Wallis test was adopted to compare feature distributions. The Benjamini-Hochberg method, which controls the false discovery rate, was implemented as a multiple-testing correction to Kruskal-Wallis and was applied per-feature for each trial to control for multiple comparisons across acquisition protocols, with FDR set by $\alpha = 0.05$. ROC-AUC and PR-AUC were unchanged when sweeping over a range of $\alpha = 0.01$ to $\alpha = 0.1$. Adjusted p-values were calculated using this method with the multipletests function from the Python statsmodels library.

We used Wilcoxon signed-rank tests to compare performance metrics between methods in trials where all models were trained. All confidence intervals are 95% confidence intervals, calculated across trials. Confidence intervals for continuous variables were calculated using the $t$-distribution and confidence intervals for discrete variables (e.g. numbers of selected features) were calculated using the Poisson distribution. These statistical tests were all conducted in Python using the SciPy library. The predictive performance of ML models was evaluated using the area under the receiver operating curve (ROC-AUC), the area under the precision-recall curve (PR-AUC), weighted accuracy, sensitivity, and specificity, all of which were calculated using the Python library scikit-learn. The Delong test was used to compare ROC-AUC in the main text and was conducted in Python using the MLstatkit library. Delong could not be performed for one trial due to <2 samples of a class being present in the test set. The Wilcoxon signed-rank test was secondarily used to compare ROC-AUC in Table 3 and Supplementary Table S6 for uniformity with the rest of the metrics.

Some early-development PNs contributed multiple scans (1.96 scans per PN on average, $\sigma = 1.02$), creating some clustering within the test set of a given fold. We primarily report results at the scan level because we wish to consider whether a model using our methodology could reduce the need for repeated scans (Table 3a, Figure 5). Note that our performance metrics, AUC, sensitivity, and specificity, do not assume sample independence. Because repeated scans weight some PNs more heavily when calculating overall performance, we also report performance at the PN level (Table 3b). Reported confidence intervals are unaffected by repeated scans of a given PN because they are not calculated within a single trial.

All *p*-values reported in this study are two-tailed. Adjusted *p*-values for the Delong test and all other classifier performance metrics were calculated with Holm-Bonferroni, which controls the family-wise error rate, using the multipletests function from the Python statsmodels library to control for multiple comparisons across the three harmonization methods.

# Data Availability

The imaging datasets used in this study are currently restricted from public release due to data privacy laws and institutional review board policy. Anonymized radiomic feature data, acquisition protocols, and diagnostic data which were analyzed in this study are available from the corresponding author on reasonable request.

# Code Availability

The code used to generate the primary results of this study is available at https://github.com/chuchthausen/HarmonizationLungRadiomics. This version of the manuscript used code with the commit tag 165be17. The Optimized Nested ComBat code by Hannah Horng (MIT license, 2022) used in this study with a minor modification is available with the commit tag b090546. The original Optimized Nested ComBat code is available with examples at https://github.com/hannah-horng/opnested-combat.

# Funding

This research was supported by the NCI Cancer Center Support Grant 5P30CA044579-27, the University of Virginia Department of Physics Bostrom Scholarships, and the NSF graduate research fellowship program, Grant No. 2141064.

# Acknowledgements

Jonathan Colen acknowledges support from the Hampton Roads Biomedical Research Consortium as part of the effort associated with the Old Dominion University-Thomas Jefferson National Accelerator Facility Joint Institute on Advanced Computing for Environmental Studies.

This material is based upon work supported in part by the National Science Foundation Graduate Research Fellowship Program under Grant No. 2141064. Any opinions, findings, and conclusions or recommendations expressed in this material are those of the authors and do not necessarily reflect the views of the National Science Foundation.

# Author Contributions

C.H. contributed to conceptualization and methods development, completed segmentations and feature extraction, curated the datasets, performed all experiments, and wrote the main manuscript text. M.S. contributed to methods development and completed segmentations and feature extraction. G.L.A.d.S.

completed segmentations and feature extraction and reviewed and edited the manuscript. J.L. sourced the data. E.J. secured funding. J.C. contributed to conceptualization and methods development, provided guidance on data science and machine learning, and reviewed and edited the manuscript. K.W. conceptualized the study, secured funding, reviewed literature, reviewed and edited the manuscript, and provided overall guidance for this work.

# Competing Interests

The authors declare no competing interests.

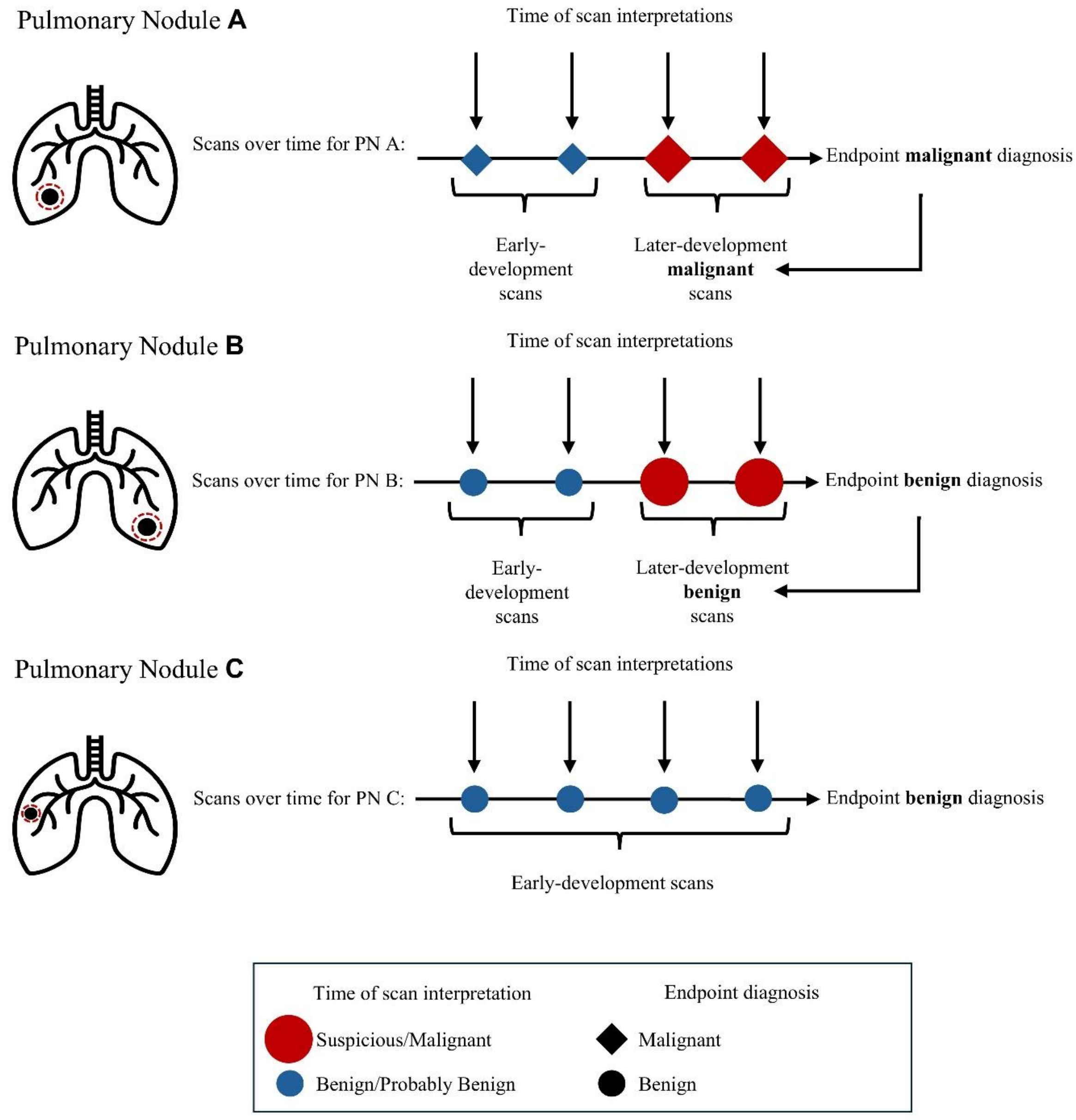

**Figure S1. Scan classification for three representative PNs.** In A and B, endpoint diagnosis was confirmed using another diagnostic method after all scans were acquired. A) A representative PN with some scans classified into the early-development dataset and some scans classified into the augmentation malignant dataset. B) A representative PN with some scans classified into the early-development dataset and some scans classified into the augmentation benign dataset. C) A representative PN with all scans classified into the early-development dataset (unconfirmed benign in Fig. 2).

**Table S1. Instances of the acquisition parameters of interest in the three datasets.** Acquisition parameters which were controlled for in this study are placed at the top of this table, and those not controlled for in this study are placed below the divider.

| Parameter | Protocol | Early-Development Occurrence (# Scans) | Augmentation Malignant Occurrence (# Scans) | Augmentation Benign Occurrence (# Scans) |
|---|---|---|---|---|
| CE | Non-CE | 80 | 57 | 18 |
| | CE | 26 | 132 | 18 |
| Focal Spots | 1.2 | 25 | 90 | 18 |
| | 0.7 | 54 | 81 | 10 |
| | 0.8 | 8 | 11 | 7 |
| | 0.9 | 19 | 7 | 1 |
| KVP | 120 | 98 | 171 | 30 |
| | 100 | 8 | 17 | 6 |
| Scanner Manufacturer | SIEMENS | 39 | 28 | 22 |
| | GE MEDICAL SYSTEMS | 67 | 161 | 14 |
| Scanner Model Name | SOMATOM Definition AS | 25 | 13 | 8 |
| | SOMATOM Force | 8 | 11 | 7 |
| | Revolution HD | 6 | 15 | 1 |
| | LightSpeed VCT | 31 | 51 | 5 |
| | Discovery CT750 HD | 30 | 75 | 8 |
| | SOMATOM Definition Flash | 6 | 3 | 1 |
| | Sensation 64 | 0 | 0 | 6 |
| Convolution Kernel | 0 | 0 | 0 | 6 |
| | BODY FILTER | 67 | 161 | 14 |
| | FLAT | 33 | 25 | 15 |
| | WEDGE_2 | 2 | 0 | 1 |
| | ['Br59f', '2'] | 19 | 6 | 1 |
| | ['I30f', '2'] | 1 | 6 | 1 |
| | LUNG | 54 | 39 | 14 |
| | ['I70f', '2'] | 7 | 2 | 5 |
| | I70f | 0 | 0 | 1 |
| | ['Bl64d', '2'] | 8 | 9 | 7 |
| | ['I70f', '3'] | 1 | 0 | 1 |
| | ['I30f', '3'] | 1 | 0 | 1 |
| | B60f | 0 | 0 | 5 |
| Data Collection Diameter | 500 | 106 | 189 | 36 |
| Slice Thickness | 1.25 | 67 | 161 | 14 |
| | 3 | 0 | 0 | 5 |
| | 5 | 0 | 0 | 1 |
| | 1.5 | 37 | 28 | 16 |

**Table S2. Frequency of net acquisition dependence for radiomic features in training sets with 100% augmentation.** Frequencies are calculated over all features and trials (n = 107 features × 50 trials = 5350).

| | Collective Harmonization (%) | | Covariate Harmonization (%) | | Separate Harmonization (%) | |
|---|---|---|---|---|---|---|
| | Augmentation Malignant | Augmentation Benign | Augmentation Malignant | Augmentation Benign | Augmentation Malignant | Augmentation Benign |
| Shape Elongation | 0 | 0 | 0 | 0 | 0 | 0 |
| Shape Flatness | 0 | 0 | 0 | 0 | 0 | 0 |
| Shape Least Axis Length | 100 | 100 | 0 | 0 | 0 | 0 |
| Shape Major Axis Length | 98 | 100 | 8 | 44 | 0 | 0 |
| Shape Maximum 2D Diameter Column | 100 | 100 | 2 | 34 | 0 | 0 |
| Shape Maximum 2D Diameter Row | 98 | 100 | 4 | 46 | 0 | 0 |
| Shape Maximum 2D Diameter Slice | 100 | 100 | 0 | 38 | 0 | 0 |
| Shape Maximum 3D Diameter | 100 | 100 | 4 | 44 | 0 | 0 |
| Shape Mesh Volume | 100 | 100 | 100 | 46 | 0 | 0 |
| Shape Minor Axis Length | 100 | 100 | 0 | 36 | 0 | 0 |
| Shape Sphericity | 100 | 72 | 0 | 2 | 0 | 0 |
| Shape Surface Area | 100 | 100 | 48 | 94 | 0 | 0 |
| Shape Surface Volume Ratio | 98 | 36 | 0 | 0 | 0 | 0 |
| Shape Voxel Volume | 100 | 100 | 100 | 46 | 0 | 0 |
| First Order 10 Percentile | 100 | 0 | 0 | 0 | 0 | 0 |
| First Order 90 Percentile | 0 | 30 | 4 | 58 | 0 | 0 |
| First Order Energy | 100 | 100 | 46 | 86 | 0 | 0 |
| First Order Entropy | 90 | 6 | 78 | 48 | 0 | 0 |
| First Order Interquartile Range | 54 | 0 | 0 | 0 | 0 | 0 |
| First Order Kurtosis | 22 | 0 | 24 | 0 | 0 | 0 |
| First Order Maximum | 2 | 0 | 2 | 10 | 0 | 0 |
| First Order Mean | 52 | 28 | 2 | 56 | 0 | 0 |
| First Order Mean Absolute Deviation | 90 | 0 | 14 | 0 | 0 | 0 |
| First Order Median | 92 | 0 | 12 | 0 | 0 | 0 |
| First Order Minimum | 100 | 100 | 98 | 10 | 100 | 0 |
| First Order Range | 10 | 0 | 2 | 4 | 0 | 0 |
| First Order Robust Mean Absolute Deviation | 48 | 8 | 0 | 0 | 0 | 0 |
| First Order Root Mean Squared | 100 | 0 | 16 | 0 | 0 | 0 |
| First Order Skewness | 98 | 8 | 0 | 0 | 0 | 0 |
| First Order Total Energy | 100 | 100 | 46 | 86 | 0 | 0 |
| First Order Uniformity | 56 | 6 | 76 | 56 | 0 | 0 |
| First Order Variance | 36 | 74 | 6 | 80 | 0 | 0 |
| GLCM Autocorrelation | 88 | 0 | 6 | 0 | 0 | 0 |
| GLCM Cluster Prominence | 4 | 16 | 0 | 52 | 0 | 0 |
| GLCM Cluster Shade | 28 | 4 | 2 | 0 | 0 | 0 |
| GLCM Cluster Tendency | 40 | 18 | 0 | 36 | 0 | 0 |
| GLCM Contrast | 52 | 56 | 48 | 58 | 0 | 0 |
| GLCM Correlation | 0 | 100 | 72 | 12 | 0 | 0 |
| GLCM Difference Average | 62 | 44 | 28 | 46 | 0 | 0 |
| GLCM Difference Entropy | 98 | 8 | 84 | 56 | 0 | 0 |
| GLCM Difference Variance | 0 | 48 | 28 | 72 | 0 | 0 |
| GLCM Inverse Difference | 72 | 98 | 10 | 58 | 0 | 0 |
| GLCM Inverse Difference Moment | 72 | 100 | 36 | 94 | 0 | 0 |
| GLCM Inverse Difference Moment Normalized | 98 | 26 | 0 | 22 | 0 | 0 |
| GLCM Inverse Difference Normalized | 98 | 26 | 0 | 28 | 0 | 0 |
| GLCM Informational Measure of Correlation 1 | 100 | 24 | 92 | 0 | 0 | 0 |
| GLCM Informational Measure of Correlation 2 | 98 | 0 | 0 | 0 | 0 | 0 |
| GLCM Inverse Variance | 74 | 86 | 2 | 72 | 0 | 0 |
| GLCM Joint Average | 80 | 0 | 14 | 0 | 0 | 0 |
| GLCM Joint Energy | 100 | 8 | 46 | 60 | 0 | 0 |
| GLCM Joint Entropy | 100 | 44 | 98 | 0 | 0 | 0 |
| GLCM Maximum Correlation Correlation | 100 | 0 | 100 | 0 | 0 | 0 |
| GLCM Maximum Probability | 88 | 52 | 46 | 82 | 0 | 0 |
| GLCM Sum Average | 80 | 0 | 14 | 0 | 0 | 0 |
| GLCM Sum Entropy | 96 | 88 | 22 | 0 | 0 | 0 |
| GLCM Sum Squares | 70 | 30 | 26 | 54 | 0 | 0 |
| GLDM Dependence Entropy | 100 | 100 | 0 | 32 | 0 | 0 |
| GLDM Dependence Non-uniformity | 100 | 98 | 4 | 0 | 0 | 0 |
| GLDM Dependence Non-uniformity Normalized | 98 | 80 | 0 | 34 | 0 | 0 |
| GLDM Dependence Variance | 100 | 100 | 100 | 100 | 0 | 0 |
| GLDM Gray Level Non-uniformity | 100 | 98 | 100 | 62 | 100 | 0 |
| GLDM Gray Level Variance | 36 | 74 | 6 | 80 | 0 | 0 |
| GLDM High Gray Level Emphasis | 80 | 0 | 8 | 0 | 0 | 0 |

| GLDM Large Dependence Emphasis | 100 | 100 | 100 | 100 | 0 | 0 |
|---|---|---|---|---|---|---|
| GLDM Large Dependence High Gray Level Emphasis | 96 | 100 | 100 | 100 | 0 | 0 |
| GLDM Large Dependence Low Gray Level Emphasis | 8 | 98 | 56 | 100 | 0 | 0 |
| GLDM Low Gray Level Emphasis | 96 | 2 | 20 | 100 | 0 | 0 |
| GLDM Small Dependence Emphasis | 90 | 40 | 0 | 34 | 0 | 0 |
| GLDM Small Dependence High Gray Level Emphasis | 66 | 98 | 100 | 36 | 0 | 0 |
| GLDM Small Dependence Low Gray Level Emphasis | 100 | 12 | 34 | 12 | 0 | 0 |
| GLRLM Gray Level Non-uniformity | 100 | 76 | 100 | 28 | 100 | 0 |
| GLRLM Gray Level Non-uniformity Normalized | 62 | 4 | 58 | 52 | 0 | 0 |
| GLRLM Gray Level Variance | 42 | 74 | 8 | 80 | 0 | 0 |
| GLRLM High Gray Level Run Emphasis | 90 | 0 | 14 | 0 | 0 | 0 |
| GLRLM Long Run Emphasis | 100 | 100 | 100 | 100 | 0 | 0 |
| GLRLM Long Run High Gray Level Emphasis | 20 | 94 | 0 | 0 | 0 | 0 |
| GLRLM Long Run Low Gray Level Emphasis | 94 | 2 | 18 | 100 | 0 | 0 |
| GLRLM Low Gray Level Run Emphasis | 96 | 0 | 20 | 100 | 0 | 0 |
| GLRLM Run Entropy | 78 | 84 | 0 | 4 | 0 | 0 |
| GLRLM Run Length Non-uniformity | 100 | 100 | 90 | 0 | 0 | 0 |
| GLRLM Run Length Non-uniformity Normalized | 100 | 100 | 64 | 58 | 0 | 0 |
| GLRLM Run Percentage | 100 | 100 | 100 | 74 | 0 | 0 |
| GLRLM Run Variance | 100 | 100 | 100 | 100 | 0 | 0 |
| GLRLM Short Run Emphasis | 100 | 100 | 86 | 72 | 0 | 0 |
| GLRLM Short Run High Gray Level Emphasis | 96 | 0 | 52 | 0 | 0 | 0 |
| GLRLM Short Run Low Gray Level Emphasis | 96 | 0 | 20 | 100 | 0 | 0 |
| GLSZM Gray Level Non-uniformity | 100 | 100 | 36 | 10 | 0 | 0 |
| GLSZM Gray Level Non-uniformity Normalized | 98 | 52 | 24 | 30 | 0 | 0 |
| GLSZM Gray Level Variance | 0 | 74 | 0 | 90 | 0 | 0 |
| GLSZM High Gray Level Zone Emphasis | 100 | 0 | 72 | 0 | 0 | 0 |
| GLSZM Large Area Emphasis | 100 | 100 | 100 | 100 | 100 | 0 |
| GLSZM Large Area High Gray Level Emphasis | 100 | 100 | 100 | 100 | 100 | 0 |
| GLSZM Large Area Low Gray Level Emphasis | 100 | 100 | 100 | 100 | 100 | 0 |
| GLSZM Low Gray Level Zone Emphasis | 100 | 0 | 62 | 98 | 0 | 0 |
| GLSZM Size Zone Non-uniformity | 100 | 100 | 18 | 0 | 0 | 0 |
| GLSZM Size Zone Non-uniformity Normalized | 84 | 4 | 10 | 30 | 0 | 0 |
| GLSZM Small Area Emphasis | 84 | 4 | 8 | 30 | 0 | 0 |
| GLSZM Small Area High Gray Level Emphasis | 70 | 0 | 64 | 0 | 0 | 0 |
| GLSZM Small Area Low Gray Level Emphasis | 100 | 0 | 54 | 78 | 0 | 0 |
| GLSZM Zone Entropy | 100 | 14 | 34 | 0 | 0 | 0 |
| GLSZM Zone Percentage | 94 | 18 | 0 | 34 | 0 | 0 |
| GLSZM Zone Variance | 100 | 100 | 100 | 100 | 100 | 0 |
| NGTDM Busyness | 12 | 94 | 92 | 100 | 0 | 0 |
| NGTDM Coarseness | 100 | 100 | 70 | 38 | 0 | 0 |
| NGTDM Complexity | 100 | 74 | 100 | 64 | 0 | 0 |
| NGTDM Contrast | 100 | 98 | 98 | 82 | 0 | 0 |
| NGTDM Strength | 100 | 100 | 100 | 50 | 0 | 0 |

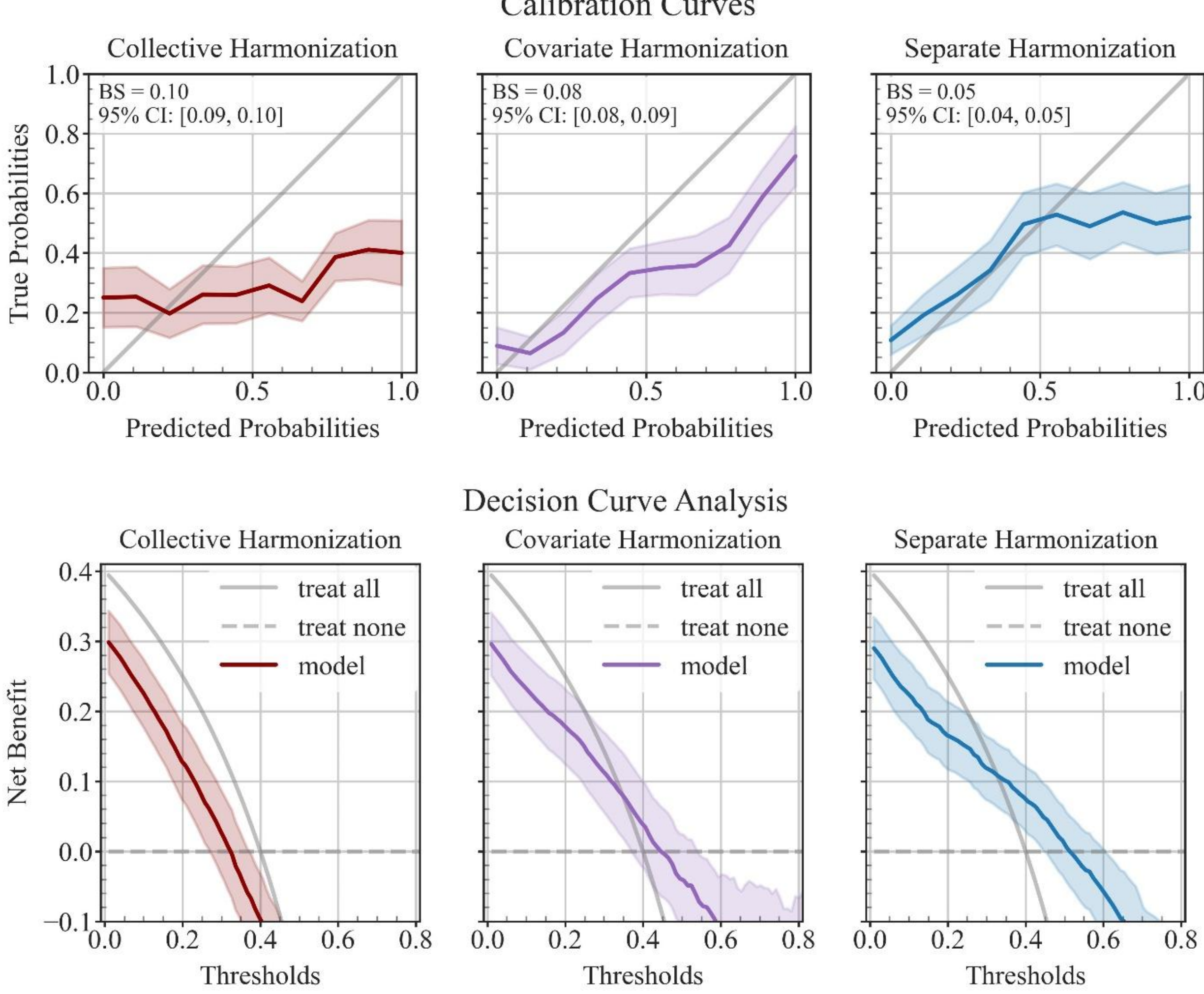

**Figure S2. Calibration and decision curve analysis for the three harmonization models.** The curves shown are calibrated via isotonic regression. The curves are averaged over 50 trials and the shaded regions indicate the 95% confidence interval. Brier scores (BS) for the calibration curves (top) are displayed in the upper left. For decision curves (bottom), note that the "treat none" strategy was used for these test samples when they were encountered in the clinic.

**Table S3. Frequency of selection for acquisition-independent radiomic features in training sets with 100% augmentation.** Features not selected after any harmonization method (0% frequency) are not shown. Features that remained acquisition-dependent after a given harmonization method, therefore not seen by the SVM, are marked by X. Frequencies are calculated over all features and trials (n = 107 features × 50 trials = 5350).

| | Collective Harmonization (%) | Covariate Harmonization (%) | Separate Harmonization (%) |
|---|---|---|---|
| Shape Elongation | 66 | 50 | 6 |
| Shape Flatness | 100 | 94 | 98 |
| Shape Least Axis Length | X | 48 | 10 |
| Shape Maximum 2D Diameter Column | X | 34 | 0 |
| Shape Maximum 2D Diameter Row | X | 48 | 94 |
| Shape Maximum 2D Diameter Slice | X | 52 | 0 |
| Shape Minor Axis Length | X | 60 | 0 |
| Shape Sphericity | X | 98 | 100 |
| Shape Surface Volume Ratio | 2 | 6 | 86 |
| First Order 10 Percentile | 0 | 8 | 0 |
| First Order 90 Percentile | 38 | 12 | 0 |
| First Order Energy | X | 6 | 0 |
| First Order Entropy | 6 | X | 24 |
| First Order Interquartile Range | 14 | 28 | 52 |
| First Order Kurtosis | 44 | 70 | 98 |
| First Order Maximum | 14 | 36 | 38 |
| First Order Mean | 4 | 6 | 0 |
| First Order Range | 72 | 8 | 60 |
| First Order Robust Mean Absolute Deviation | 8 | 40 | 0 |
| First Order Root Mean Squared | 0 | 20 | 2 |
| First Order Skewness | 2 | 4 | 0 |
| First Order Total Energy | X | 6 | 0 |
| First Order Uniformity | 10 | 2 | 0 |
| First Order Variance | X | 10 | 0 |
| GLCM Autocorrelation | 0 | 10 | 0 |
| GLCM Cluster Prominence | 48 | 4 | 0 |
| GLCM Cluster Shade | 70 | 98 | 100 |
| GLCM Cluster Tendency | 36 | 12 | 2 |
| GLCM Contrast | 16 | X | 6 |
| GLCM Correlation | X | 16 | 20 |
| GLCM Difference Average | 18 | 28 | 18 |
| GLCM Difference Entropy | X | X | 16 |
| GLCM Difference Variance | 32 | 4 | 14 |
| GLCM Inverse Difference | 2 | 24 | 0 |

| GLCM Inverse Difference Moment | X | 2 | 0 |
|---|---|---|---|
| GLCM Inverse Difference Moment Normalized | 2 | 56 | 6 |
| GLCM Inverse Difference Normalized | X | 30 | 0 |
| GLCM Informational Measure of Correlation 1 | X | 8 | 0 |
| GLCM Informational Measure of Correlation 2 | 2 | 82 | 100 |
| GLCM Inverse Variance | 12 | 4 | 60 |
| GLCM Joint Average | 4 | 0 | 0 |
| GLCM Joint Energy | X | 18 | 10 |
| GLCM Maximum Correlation Correlation | 0 | 0 | 18 |
| GLCM Maximum Probability | X | 6 | 6 |
| GLCM Sum Average | 2 | 0 | 0 |
| GLCM Sum Entropy | 4 | 60 | 10 |
| GLCM Sum Squares | 4 | 2 | 0 |
| GLDM Dependence Entropy | X | 60 | 0 |
| GLDM Dependence Non-uniformity | X | 6 | 0 |
| GLDM Dependence Non-uniformity Normalized | 2 | 60 | 6 |
| GLDM Dependence Variance | X | X | 2 |
| GLDM Gray Level Variance | X | 4 | 0 |
| GLDM High Gray Level Emphasis | 12 | 4 | 0 |
| GLDM Large Dependence High Gray Level Emphasis | X | X | 100 |
| GLDM Large Dependence Low Gray Level Emphasis | X | X | 86 |
| GLDM Small Dependence Emphasis | 6 | 10 | 0 |
| GLDM Small Dependence High Gray Level Emphasis | X | X | 28 |
| GLDM Small Dependence Low Gray Level Emphasis | X | 38 | 0 |
| GLRLM Gray Level Non-uniformity Normalized | 22 | X | 0 |
| GLRLM High Gray Level Run Emphasis | 0 | 30 | 0 |
| GLRLM Long Run High Gray Level Emphasis | 6 | 100 | 50 |
| GLRLM Long Run Low Gray Level Emphasis | 2 | X | 0 |
| GLRLM Run Entropy | 12 | 4 | 0 |
| GLRLM Run Length Non-uniformity | X | 4 | 0 |
| GLRLM Run Length Non-uniformity Normalized | X | 8 | 0 |
| GLRLM Short Run High Gray Level Emphasis | 2 | 14 | 0 |
| GLSZM Gray Level Non-uniformity | X | 8 | 0 |
| GLSZM Gray Level Non-uniformity Normalized | 2 | 30 | 34 |
| GLSZM Gray Level Variance | 26 | 8 | 28 |
| GLSZM High Gray Level Zone Emphasis | 0 | 2 | 0 |
| GLSZM Low Gray Level Zone Emphasis | 0 | X | 6 |
| GLSZM Size Zone Non-uniformity | X | 8 | 0 |
| GLSZM Size Zone Non-uniformity Normalized | 12 | 18 | 68 |
| GLSZM Small Area Emphasis | X | 30 | 62 |
| GLSZM Small Area High Gray Level Emphasis | 28 | 36 | 0 |
| GLSZM Small Area Low Gray Level Emphasis | 0 | 10 | 100 |
| NGTDM Busyness | 4 | X | 0 |
| NGTDM Coarseness | X | 4 | 0 |
| NGTDM Complexity | X | X | 4 |
| NGTDM Contrast | X | X | 16 |
| NGTDM Strength | X | X | 100 |

**Table S4. Pearson correlation coefficients demonstrate feature selection stability across trials.** Pearson correlation was calculated for Boolean arrays indicating feature selection (1=selected, 0=not selected), determined by the L1-regularized SVM, per pair of trials for features which were acquisition-independent in all trials, the number of which varied per harmonization method. Magnitudes of the Pearson Correlation Coefficient (PCC) near 1 indicate the model gives similar importance to features across paired trials. The correlation was calculated once per nonidentical trial pairing. The numbers of feature pairings were, collective harmonization: n=136, covariate harmonization: n=1225, separate harmonization: n=1225. Collective harmonization has a different number of pairings because the test frequently failed when the input coefficient array(s) were constant due to collective harmonization's small number of compared features. Correlation of radiomic feature values within a trial was also assessed using the PCC over n=50 trials. Features were considered correlated when PCC > 0.7. Correlations between radiomics feature values themselves may lower stability scores, as models may select different features across trials to access the same information.

| | **Collective Harmonization** | **Covariate Harmonization** | **Separate Harmonization** |
|---|---|---|---|
| Selection PCC Mean | 1.0 | 0.48 | 0.69 |
| Selection PCC Median | 1.0 | 0.53 | 0.69 |
| Selection PCC Standard Deviation | 0.0 | 0.27 | 0.09 |
| Average Number of Correlated Feature Pairs | 0.04<br>[0.0-0.14] | 2.2<br>[1.8-2.6] | 518.5<br>[512.2-524.9] |
| Trials with at least one correlated feature | 2 | 50 | 50 |
| Acquisition-independent features | 2 | 10 | 100 |

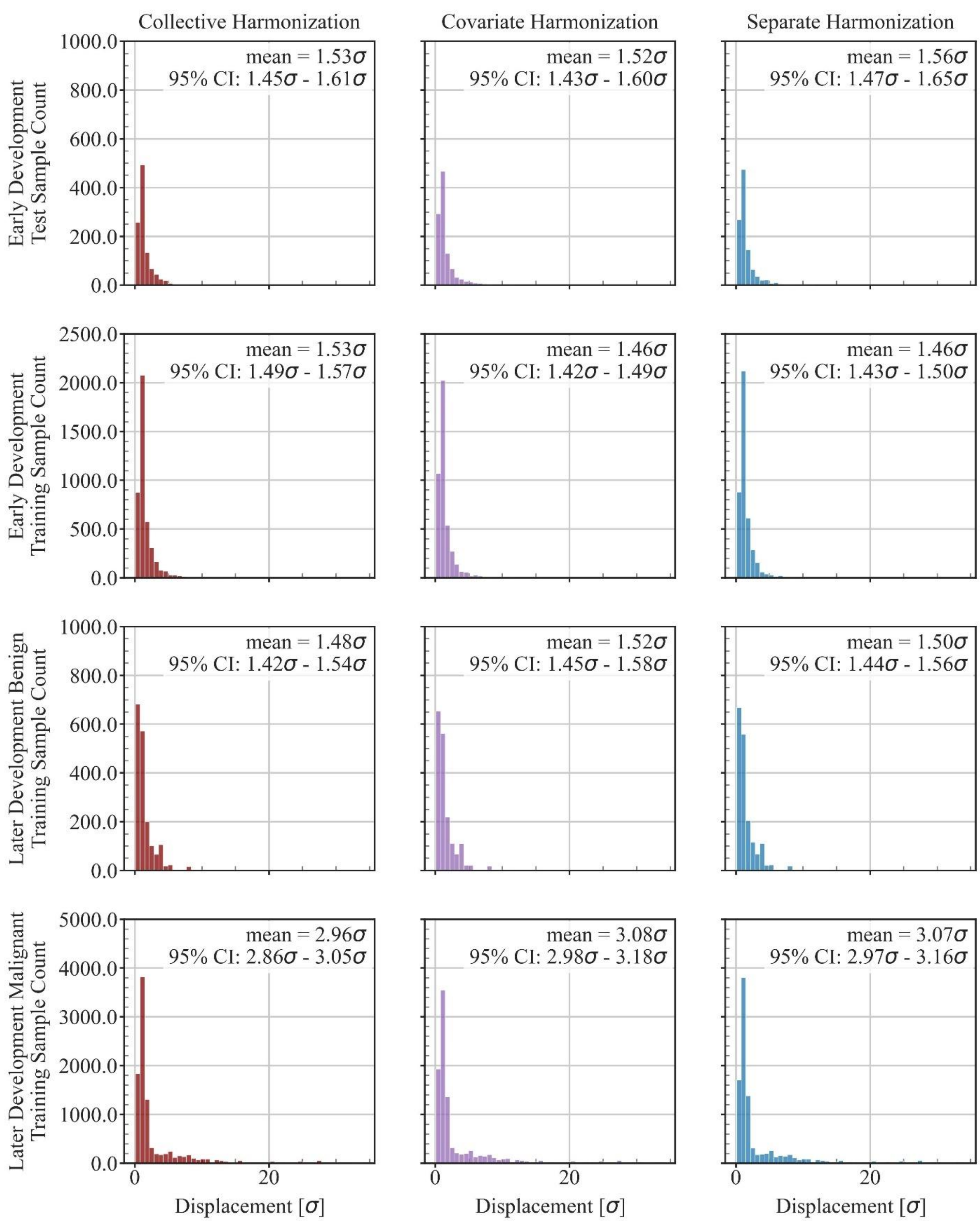

**Figure S3. Displacement of samples in feature space produced by harmonization.** All unharmonized samples and harmonized samples for all trials were Z-score normalized in a common feature space. Sample displacements were then calculated in the space of features which were acquisition-independent for each trial. Displacements are reported in units of standard deviation.

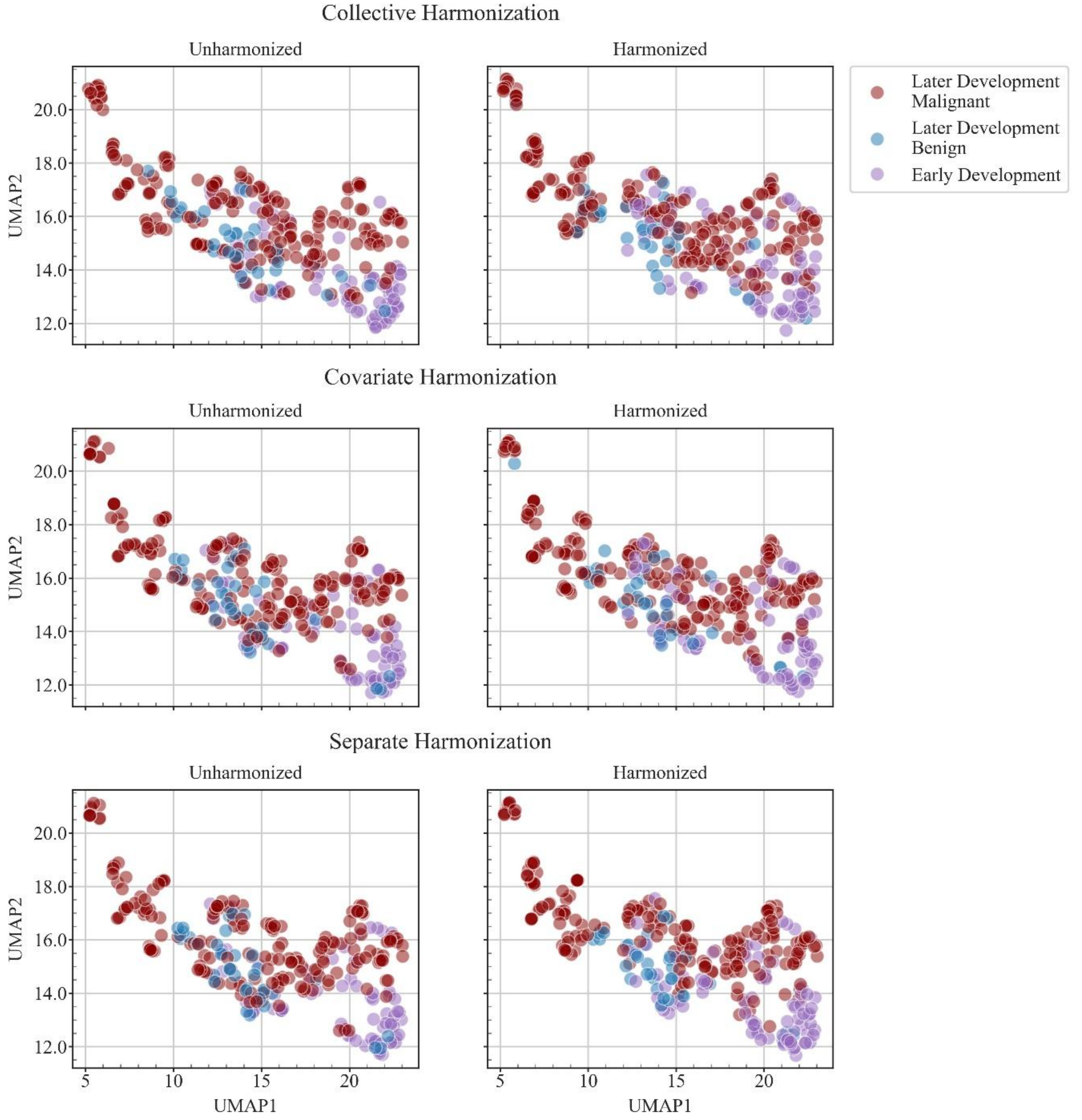

**Figure S4. UMAPs of a training set before and after harmonization for a representative training set.** All training samples, unharmonized and all harmonization versions, are projected into a common two-dimensional feature space. Only features which are acquisition-independent after harmonization for the respective harmonization method are displayed in each row. Because different features are acquisition independent for different methods, the unharmonized samples are displayed as shifted in the common feature space prior to harmonization.

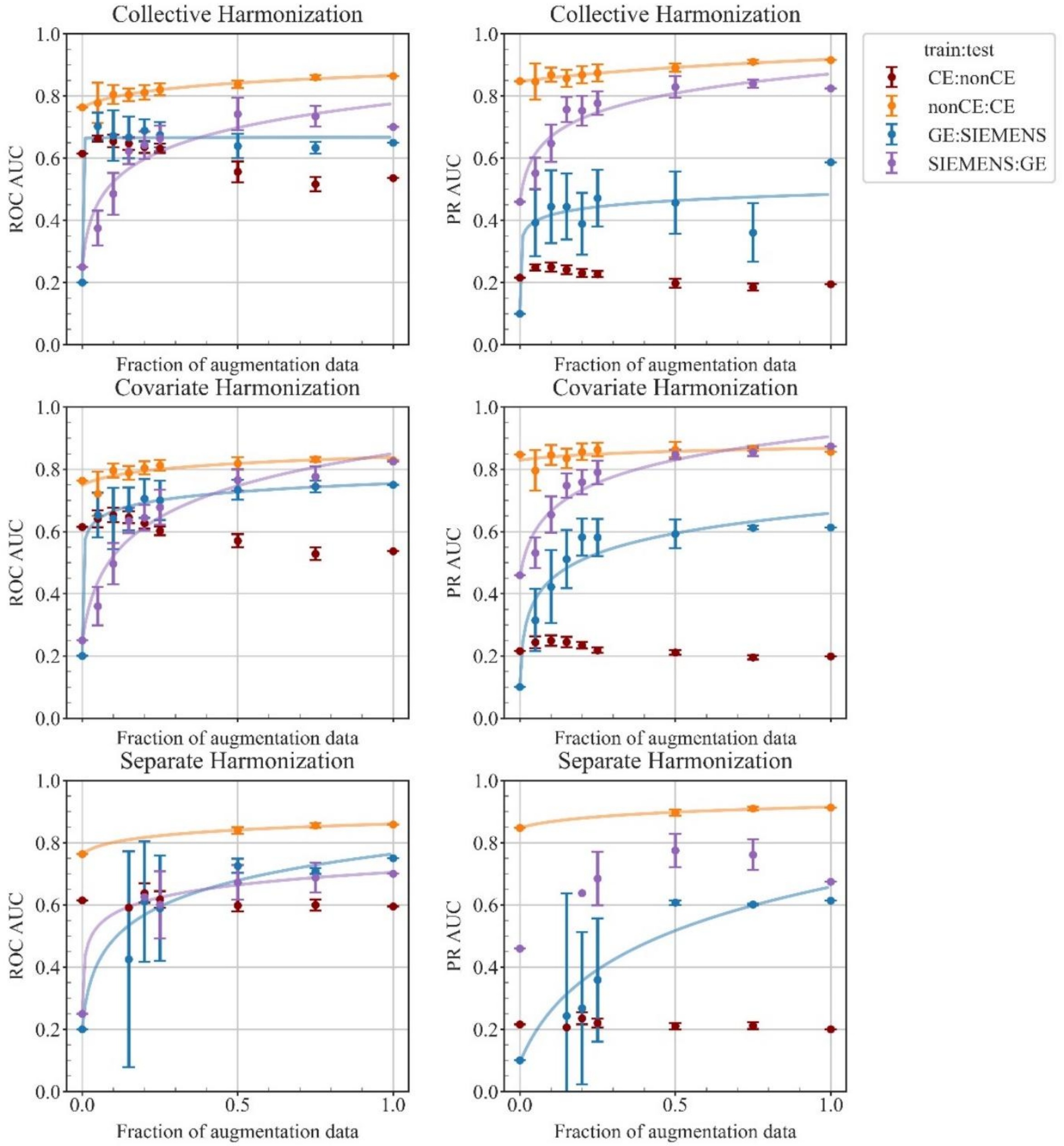

| | $R^2$, +log Collective Harmonization | | $R^2$, +log Covariate Harmonization | | $R^2$, +log Separate Harmonization | |
|---|---|---|---|---|---|---|
| **Train : Test** | **ROC-AUC** | **PR-AUC** | **ROC-AUC** | **PR-AUC** | **ROC-AUC** | **PR-AUC** |
| CE : Non-CE | - | - | - | - | - | - |
| Non-CE : CE | 0.31 | 0.18 | 0.18 | 0.04 | 0.91 | 0.86 |
| GE : SIEMENS | 0.83 | 0.39 | 0.75 | 0.63 | 0.83 | 0.77 |
| SIEMENS : GE | 0.79 | 0.76 | 0.82 | 0.79 | 0.87 | **-** |

**Figure S5. Test cases for pipeline performance on unseen batch effects.** Performance is calculated at the scan level. The learning curves for models trained on early-development scans acquired with one manufacturer or CE status and tested on the held-out set, obtained by varying the fraction of available augmentation data included in the training set. Multiple trials for 5%-75% augmentation were obtained by varying the random splitting of the available

augmentation data into augmentation fractions. 20 trials were attempted for each point, but some trials failed at low augmentation due to rare acquisition parameters. Error bars show the 95% confidence interval. Per curve, the data is not harmonized for the acquisition parameter by which the split is made, and only augmentation data corresponding to the training batch is included, creating an unseen batch effect. The leftmost point in each plot represents the baseline un-augmented model. Higher baseline performance compared to our primary pipeline is likely due to removing, necessarily, the constraint isolating scans from a patient within a fold. Performance increases with augmentation fraction for three of four test cases. Data from CE scans does not generalize well to non-CE scans. Gains from augmentation were not stable for PR-AUC when training on separately harmonized Siemens data and testing on GE. Where data could be fitted to a positive logarithmic curve, $R^2$ fits are given in the associated table.

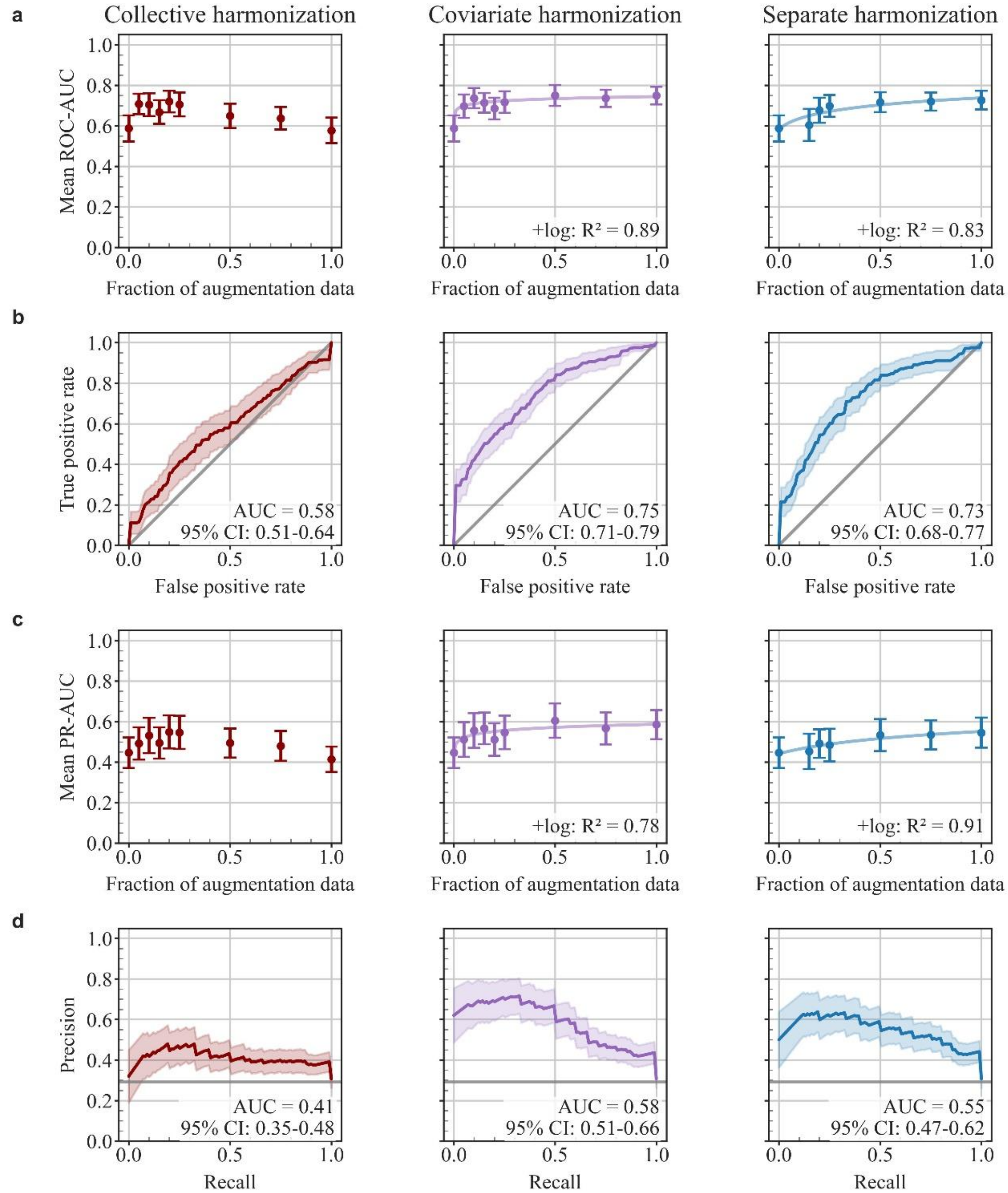

**Figure S6. The average performance of the three harmonization models with hyperparameters tuned via nested cross-validation.** a. (c.) The ROC-AUC (PR-AUC) learning curves for hyperparameter-tuned models with augmented training sets obtained by varying the fraction of augmentation data included in the training set. The leftmost point in each plot represents the baseline un-augmented model. Error bars show the 95% confidence interval over trials where a model could be trained. Trial numbers *n* for each point, from left to right—collective harmonization: *n*=50 for all, covariate harmonization: *n*=50 for all, separate harmonization: *n*=[50, 10, 30, 45, 50, 50, 50]. Separate harmonization failed for 5% and 10% augmentation due to small sample size, so these points are omitted. b. (d.) Receiver operating (precision-recall) curves for hyperparameter-tuned models trained using 100% of augmentation data, averaged over all trials, n=50 for each. Shaded regions show the 95% confidence interval. Gray lines indicate chance-level performance.

**Table S5. The contents of the early-development dataset.** PNs within this dataset, classified primarily by physician time-of-scan diagnosis, may be further classified by their endpoint diagnosis and endpoint diagnosis method. See Methods for a description of how endpoint diagnosis is determined and for the stability criteria used in this study for classifying some PNs.

| Endpoint Diagnosis | Endpoint Diagnosis Method | Number of Patients | Number of PNs | Number of Scans |
|---|---|---|---|---|
| Malignant | PET or biopsy confirmation | 60 | 61 | 220 |
| Benign | PET or biopsy confirmation | 25 | 35 | 56 |
| Benign | Lung-RADs and stability criteria | 15 | 24 | 55 |

**Table S6. Performance metrics for the three harmonization models evaluated on only test scans with confirmed labels.** Results are reported at the scan level. Presumed-benign PNs are excluded. 100% augmentation data are included. Weighted accuracy uses balanced weighting by class (endpoint benign or malignant diagnosis). *Significant (Wilcoxon, adjusted p<0.05) improvement over collective harmonization. Due to small sample size, only improvement to PR-AUC was significant.

| | Collective Harmonization | | Covariate Harmonization | | Separate Harmonization | |
|---|---|---|---|---|---|---|
| | Mean, % | 95% CI | Mean, % | 95% CI | Mean, % | 95% CI |
| ROC AUC | 62.5 | [54.2, 70.8] | 75.2 | [66.9, 83.6] | 71.3 | [62.8, 79.8] |
| PR AUC | 73.4 | [65.8, 81.0] | 82.2* | [75.8, 88.6] | 78.3* | [71.1, 85.6] |
| Weighted accuracy | 61.5 | [56.4, 66.6] | 62 | [56.1, 67.9] | 58.1 | [51.8, 64.4] |
| Specificity at 90% sensitivity | 30.7 | [19.5, 41.9] | 49 | [36.6, 61.5] | 45.6 | [33.3, 58.0] |
| Sensitivity at 30% specificity | 79.9 | [71.4, 88.4] | 85.2 | [76.5, 94.0] | 82.4 | [73.4, 91.4] |
| Sensitivity at 90% specificity | 35.1 | [23.7, 46.5] | 51.3 | [39.1, 63.4] | 44.9 | [31.8, 57.9] |

**Table S7. Acquisition parameter-based training scan exclusion.** Mean numbers of training scans excluded per trial due to rare instances of an acquisition parameter. All excluded scans were later-development scans. Rare focal spots was responsible for 35% of exclusions, rate KVP was responsible for 33% of exclusions, rare CE (during separate harmonization) was responsible for 24% of exclusions, and rare manufacturer was responsible for 8% of exclusions. Trial numbers *n*, from top to bottom—collective harmonization: *n*=50 for all, covariate harmonization: *n*=50 for all, separate harmonization: *n*=[50, 10, 30, 45, 50, 50, 50].

| | Collective Harmonization | | Covariate Harmonization | | Separate Harmonization | |
|---|---|---|---|---|---|---|
| Augmentation fraction (%) | Mean | 95% CI | Mean | 95% CI | Mean | 95% CI |
| 0 | 0 | [0 - 0.07] | - | - | - | - |
| 5 | 0.1 | [0.03 - 0.23] | 0.1 | [0.03 - 0.23] | - | - |
| 10 | 0.1 | [0.03 - 0.23] | 0.1 | [0.03 - 0.23] | - | - |
| 15 | 0.1 | [0.03 - 0.23] | 0.1 | [0.03 - 0.23] | 7.5 | [5.90 - 9.40] |
| 20 | 0.1 | [0.03 - 0.23] | 0.1 | [0.03 - 0.23] | 5.17 | [4.39 - 6.05] |
| 25 | 0.3 | [0.17 - 0.49] | 0.3 | [0.17 - 0.49] | 4.67 | [4.06 - 5.34] |
| 50 | 0.7 | [0.49 - 0.97] | 0.7 | [0.49 - 0.97] | 3.9 | [3.37 - 4.49] |
| 75 | 0.9 | [0.66 - 1.20] | 0.9 | [0.66 - 1.20] | 1.9 | [1.54 - 2.32] |
| 100 | 1 | [0.74 - 1.32] | 1 | [0.74 - 1.32] | 2 | [1.63 - 2.43] |